\documentclass[10pt,journal,cspaper,compsoc]{IEEEtran}

\ifCLASSOPTIONcompsoc
\usepackage[nocompress]{cite}
\else
\usepackage{cite}
\fi

\usepackage{graphicx}
\usepackage{subfigure}
\usepackage{epsfig}
\usepackage{amsmath}
\usepackage{amssymb}
\usepackage{cases}
\usepackage{hhline}
\usepackage{caption}
\usepackage{algorithm}
\usepackage{algorithmic}
\usepackage[T1]{fontenc}
\usepackage{color}

\begin{document}
\title{Security-Sensitive Task Offloading in Integrated Satellite-Terrestrial Networks}

\author{Wenjun Lan, Kongyang Chen, Jiannong Cao,~\IEEEmembership{Fellow,~IEEE}, Yikai Li, Ning Li, Qi Chen, and Yuvraj Sahni
\IEEEcompsocitemizethanks{
\IEEEcompsocthanksitem{W. Lan, K. Chen, Y. Li, N. Li, and Q. Chen are with Institute of Artificial Intelligence, Guangzhou University, China. K. Chen is also with Pazhou Lab, Guangzhou, China. E-mail: kychen@gzhu.edu.cn. (Corresponding author: K. Chen)}
\IEEEcompsocthanksitem{J. Cao is with Department of Computing, The Hong Kong Polytechnic University, Hong Kong, China.} 
\IEEEcompsocthanksitem{Y. Sahni is with Department of Building Environment and Energy Engineering, The Hong Kong Polytechnic University, Hong Kong, China.} 
}}

\IEEEtitleabstractindextext{
\begin{abstract}
With the rapid development of sixth-generation (6G) communication technology, global communication networks are moving towards the goal of comprehensive and seamless coverage. In particular, low earth orbit (LEO) satellites have become a critical component of satellite communication networks. The emergence of LEO satellites has brought about new computational resources known as the \textit{LEO satellite edge}, enabling ground users (GU) to offload computing tasks to the resource-rich LEO satellite edge. However, existing LEO satellite computational offloading solutions primarily focus on optimizing system performance, neglecting the potential issue of malicious satellite attacks during task offloading. In this paper, we propose the deployment of LEO satellite edge in an integrated satellite-terrestrial networks (ISTN) structure to support \textit{security-sensitive computing task offloading}. We model the task allocation and offloading order problem as a joint optimization problem to minimize task offloading delay, energy consumption, and the number of attacks while satisfying reliability constraints. To achieve this objective, we model the task offloading process as a Markov decision process (MDP) and propose a security-sensitive task offloading strategy optimization algorithm based on proximal policy optimization (PPO). Experimental results demonstrate that our algorithm significantly outperforms other benchmark methods in terms of performance.
\end{abstract}

}
	
\maketitle
\IEEEdisplaynontitleabstractindextext
\IEEEpeerreviewmaketitle

\section{Introduction}\label{sec:introduction}
The emerging sixth-generation (6G) communication technology is rapidly gaining momentum, with its vision being to establish a ubiquitous communication network that covers the entire globe, catering to the surging demand for intelligent devices and data traffic. Although the formal regulation and commercial deployment of 6G have yet to commence, both academia and industry have been actively engaged in research and development efforts related to 6G communication technology. 6G holds the promise of providing wireless communication services across the entire globe, supporting a wide range of Internet of Things (IoT) applications, especially in remote areas. However, the current coverage of terrestrial networks is relatively limited, primarily concentrated in major urban areas, which impedes the broader proliferation and application of communication networks\cite{8610431}.

Under the circumstances, satellite communication networks have emerged as a promising and highly potential solution\cite{7105655}. Compared to medium earth orbit (MEO) and geostationary orbit (GEO) satellites, low earth orbit (LEO) satellites are closer to the Earth, making them more suitable for supporting latency-sensitive communications on a global scale.  LEO satellites possess high capacity and wide coverage capabilities, which are currently experiencing rapid development\cite{you2021towards, giordani2020non, chen2020system, zhu2021integrated}. Companies like SpaceX and OneWeb are driving the large-scale deployment of LEO constellations, redefining network architectures to enable low latency, high capacity, and global service\cite{9473799, 8688478, 8571192}. This presents a fresh opportunity for achieving a global network. The low latency communication capabilities of LEO satellites make them an ideal choice for Internet of Things applications, while the rise of large-scale LEO constellations is fundamentally transforming network architectures to achieve the objectives of low latency, high capacity, and global coverage. This lays a solid foundation for realizing the vision of a \textit{connected world} and provides reliable communication connectivity for remote areas, bridging the digital divide.

Recent research findings strongly confirm the immense potential of satellite networks in addressing ground backhaul link failures or congestion issues and significantly enhancing the offloading capabilities of remote base stations\cite{9978929, 10018447}. Particularly, with the utilization of LEO satellites, serving as alternative wireless access nodes, they effectively support users in offloading tasks beyond the coverage range of base stations or to remote regions with poor channel conditions. This brings important inspiration for the upcoming era of 6G, where LEO satellite networks will play a crucial role in bridging the global digital divide and meeting the growing and diverse communication demands. In this emerging paradigm, satellites can form clusters or constellations, enabling coordinated communication to provide more efficient and comprehensive coverage for users on the Earth's surface. Therefore, integrated satellite-terrestrial networks (ISTN) will dominate in the coming years. The ISTN includes LEO satellites with considerable computational capabilities, which give rise to a new paradigm as the \textit{LEO satellite edge}. This new edge computing paradigm will bring higher efficiency and convenience to various application scenarios, creating more opportunities for future smart cities and the connected world. Given the limited computational resources at the edge of LEO satellites, much like most edge computing scenarios, computation offloading emerges as a potential solution capable of significantly enhancing system performance\cite{8016573}. In the scenario of ISTN, when end-user devices generate computational tasks, they can offload tasks directly to the LEO satellite edge for real-time processing, which reduces frequent data transfers between devices and the cloud, thereby reducing end-to-end communication latency and expanding service coverage.

However, in practical applications, due to the public nature of network access, the ever-changing network topology, and potential risks associated with wireless channels, ground users may encounter privacy and security risks during the data transmission process of task offloading. Furthermore, it is essential to prevent the offloaded data from being attacked by malicious satellites. Therefore, security in the satellite-terrestrial link of ISTN becomes a crucial focal point\cite{9882109, zhu2021integrated}. In task offloading, the involved data may include sensitive information such as personal identities, trade secrets, or research data. If these data are transmitted in plaintext without any encryption measures, there will undoubtedly be security risks. To ensure the security of data transmission in ISTN, it becomes crucial to employ advanced encryption techniques to protect user data transmitted over satellite-terrestrial links. Encryption technology can effectively safeguard the confidentiality and integrity of data, thus preventing unauthorized access and tampering. Even if a malicious satellite successfully intercepts encrypted data, it cannot decrypt it to obtain useful information. In addition to encryption techniques, when formulating task offloading strategies, it is crucial to consider potential security threats to ensure that the task data within the ISTN maintains a high level of information security strength and minimizes the risk of malicious satellite attacks.

Deep reinforcement learning (DRL) is widely regarded as an efficient approach for optimizing offloading strategies in complex network environments. In recent years, there has been a surge in research on satellite edge computing offloading based on DRL. In satellite edge computing, DRL is applied to optimize offloading strategies, which determine the optimal way to offload tasks between satellites and ground servers. By using DRL, the system can learn and optimize decisions continuously by receiving feedback from the environment, resulting in more efficient computation offloading solutions. This approach is particularly suitable for complex network environments as it can adapt to changing network conditions and task requirements. However, the unique nature of satellite edge computing poses some challenges to the application of DRL in practice. Firstly, the latency and bandwidth constraints of satellite communication links require the system to make decisions within limited time and resources. Secondly, satellite edge computing systems must have a high level of security, considering the security requirements of tasks and block cipher lengths to prevent unauthorized access and data leakage. Additionally, due to limited satellite resources, latency and energy consumption management is also a crucial consideration. Lastly, the instability and unreliability of satellite communication environments can lead to network disruptions and data loss, necessitating a highly reliable system. Therefore, while some studies have made some progress, achieving highly secure, low-latency, low-energy, and high-reliability computation offloading in satellite edge computing still requires overcoming many challenges.

To address the aforementioned challenges, we have considered not only random task arrivals, the high mobility of satellites, and the uncertainty of wireless channel states but also security factors such as task security level requirements and block cipher lengths. Our objective is to minimize offloading latency, energy consumption, and the number of attacks while satisfying reliability constraints. To achieve this goal, we have employed a joint optimization approach for satellite allocation and offloading orders. We have modeled this problem as a Markov decision process (MDP) for discrete offloading actions to comprehensively consider various complex factors. Finally, we have proposed a security-sensitive task offloading strategy optimization algorithm based on proximal policy optimization (PPO) to make offloading decisions adaptively. The main contributions of this paper are summarized as follows:

\begin{enumerate}
\item We propose the deployment of the ISTN structure with LEO satellite edge that takes into account factors such as random task arrivals, task security requirements, the high mobility of LEO satellites, and the variability of wireless channel conditions.  It aims to effectively support security-sensitive task offloading for ground user. We formalize the task allocation and offloading order problem as a joint optimization problem. The objective is to minimize a weighted sum of task offloading latency, energy consumption, and the number of attacks while satisfying reliability constraints.
		
\item We model the task offloading process in ISTN as an MDP and propose a security-sensitive task offloading strategy optimization algorithm based on PPO to jointly optimize task allocation decisions and offloading order and learn the optimal task offloading strategy. 

\item Extensive experiments demonstrate that our proposed algorithm outperforms greedy algorithm, round-robin algorithm, all-local algorithm, and all-offloading algorithm, and performs to some extent better than TRPO-based algorithm and A2C-based algorithm.
		
\end{enumerate}

The remaining sections of this paper are organized as follows. Section \ref{sec:Related Work} provides a review of related works. Section \ref{sec:System Model} introduces our system model and problem formulation. Section \ref{sec:Security-Sensitive Task Offloading Solution} establishes the MDP framework and describes the PPO-based algorithm. Section \ref{sec:Numerical Results} shows the experimental results. Finally, Section \ref{sec:conclusion} concludes this paper.

\section{Related Work}\label{sec:Related Work}

\subsection{Computation Offloading in Satellite Edge Computing}
With the advancement of mobile edge computing (MEC) technology\cite{8976180}, researchers have started applying it to satellite networks, which brings computational resources to the edge of satellite networks. Satellite edge computing networks, as a novel network architecture, harness the strengths of both satellite communication and ground networks. Therefore, satellite edge computing networks hold significant potential in addressing task offloading and tackling communication latency issues.

In the scenario of a multi-layer Ka/Q-band integrated satellite-ground network for the Internet of Remote Things (IoRT), Chen et al.\cite{9638998} have introduced a dynamic computation offloading solution. They have employed a DRL algorithm based on the deep double Q network (D3QN) to optimize the selection of offloading paths and resource allocation. This approach aims to maximize the number of offloaded tasks while meeting latency requirements and minimizing satellite energy consumption.
Tang et al.\cite{9344666} proposed a hybrid cloud and edge computing LEO satellite (CECLS) network with a three-layer computing architecture. This network leverages the resources of both cloud and edge servers to provide heterogeneous computing capabilities to ground users. Additionally, it employs a distributed algorithm based on the alternating direction method of multipliers (ADMM) to effectively reduce the energy consumption of ground users.
The authors of \cite{9583851} proposed an integrated satellite-aerial-terrestrial (I-SAT) network. This network aims to improve the throughput of users in the I-SAT network by jointly optimizing user association, transmit power, and drone trajectory control. To achieve this, they employed an alternating iterative algorithm based on the block descent method, which effectively enhances the throughput of users in the I-SAT network.
Zhou et al.\cite{9222519} researched task scheduling problems in space-air-ground integrated networks (SAGIN). They proposed a solution called delay-oriented IoT task scheduling (DOTS), which aims to reduce task processing latency and meet energy constraints.
Chai et al.\cite{10024305} researched a multi-task mobile edge computing system in satellite IoT. They prioritized tasks in the directed acyclic graph (DAG) based on their cost consumption and proposed a collaborative optimization algorithm called attention mechanism and proximal policy optimization (A-PPO). Simulation results demonstrate that this approach effectively reduces the long-term cost of the offloading system.
Cao et al.\cite{9978929} conducted research on a LEOS edge-assisted multi-layer multi-access edge computing system to support computation-intensive and latency-sensitive services. They achieved a low-latency and energy-efficient offloading scheme by jointly optimizing the allocation of communication and computing resources.
Cui et al.\cite{9955992} researched the delay optimization problem in the hybrid GEO-LEO satellite-assisted Internet of Things (SIoT) network. They proposed an intelligent task offloading and multidimensional resource allocation (TOMRA) algorithm to effectively reduce the system latency of IoT tasks.
Ding et al.\cite{9515574} researched a satellite-aerial integrated edge computing network (SAIECN) that combines LEO satellites and high-altitude platforms (HAPs) to provide edge computing services to ground user equipment (GUE). Their work focused on minimizing the energy consumption of SAIECN by jointly optimizing GUE association, multi-user multiple input and multiple output (MU-MIMO) precoding, task allocation, and computing resource allocation.
The authors of \cite{10032271} proposed an intelligent satellite edge computing method. This method is based on the deep deterministic policy gradient (DDPG) algorithm, which addresses the challenge of simultaneously learning discrete and continuous actions. They jointly optimize offloading decisions and resource allocation to minimize system energy consumption while satisfying latency and resource constraints.

Based on the aforementioned literature, the issue of computation offloading in satellite MEC environments is highly complex. However, most of the existing research fails to consider the potential failures that can occur during task offloading, the coverage time limitations imposed by the high-speed movement of satellites, and the aspect of security. These studies typically rely on traditional optimization methods, which restrict the full utilization of computing resources and the reasonability of edge computing strategies. To address these limitations, our research proposes a comprehensive network model that takes into account offloading latency, energy consumption, reliability, and security concerns. Then, we utilize a DRL algorithm to optimize the offloading strategies. This innovation represents significant progress and contribution to the existing body of research literature.

\subsection{Security Solutions in Satellite Network}
The unique nature of satellite networks necessitates heightened attention to security issues when communicating between ground users or base stations and satellites. This attention is not only because communication involves sensitive data but also due to satellite networks face a range of unique security challenges in their environment and conditions\cite{10209551}. In satellite networks, it is necessary to implement multi-layered security measures, including encryption, authentication, network monitoring, and attack detection, to ensure the confidentiality, integrity, and availability of communication data.

Liu et al.\cite{9359657} designed a privacy-preserving trust management (PPTM) scheme in the scenario of the SAGIN. This scheme establishes a secure and reliable communication framework for emergencies. It achieves precise trust management and robust conditional privacy-preserving while maintaining low communication overhead.
Liao et al.\cite{9837848} proposed a dual-layer Stackelberg game model aimed at addressing the physical layer security issues in the information feedback link of the Satellite-UAV integrated (SUI) network. They designed a three-stage optimal response iterative (TORI) algorithm to find the game equilibrium and encourage the participation and cooperation of UAVs through a price incentive mechanism.
Li et al.\cite{9514547} proposed an innovative architecture consisting of satellites and ground devices, incorporating blockchain technology to protect satellite networks from threats of unauthorized information access and usage, thereby enhancing communication security. This architecture achieves authentication and privacy protection of the communication network at various stages, providing users with a more reliable protection mechanism.
To enhance the reliability and security of the integrated 5G-satellite network, Lin et al.\cite{9232927} proposed a novel beamforming scheme. The objective of this scheme is to strike a balance between security and reliability, to improve system performance.
The authors of \cite{8660485} researched the design of secure multicast communication in cognitive satellite-ground networks. To enhance the security of satellite links, they proposed a method that optimizes transmission power and leverages ground interference, thereby effectively safeguarding the confidentiality of communication content and ensuring reliable information transmission.
Tang et al.\cite{9918062} proposed a blockchain-based secure federated learning framework. This framework incorporates a node security evaluation mechanism and an enhanced Byzantine fault-tolerant algorithm to optimize routing paths and minimize overall latency in the offloading process of the SAGIN architecture while satisfying security constraints. Additionally, a blockchain-based federated asynchronous advantage actor-critic (BFA3C) algorithm is employed to optimize the decision-making process.
Liao et al.\cite{9652086} proposed a secure and low-latency electromagnetic interference-aware computation offloading algorithm based on blockchain and semi-distributed learning. This algorithm combines the technologies of blockchain, space-air-ground integrated power Internet of Things (SAG-PIoT), and machine learning to achieve secure and low-latency computation offloading.
However, the model of \cite{9918062} and \cite{9652086} ignores the energy consumption and reliability of the transmission.
The authors of \cite{9651535} proposed a secure-aware computation offloading algorithm for space computing platforms. This algorithm, based on DDPG, models risk as a Poisson distribution and optimizes time, energy, and risk components. Its objective is to ensure system security while efficiently saving time and energy.  However, their study focused on static scenarios involving satellites and space stations, which differ from the dynamic scenarios discussed in this paper regarding ISTN. Additionally, their research did not consider the dynamics of satellite networks and the physical limitations of communication transmission, which may result in data transmission errors.

In summary, although there have been some efforts to maintain the security of communication links and network architecture in satellite networks, the attention given to the reliability and security issues of task offloading, a crucial step in satellite networks, has been limited. In the process of task offloading in satellite networks, it is necessary to comprehensively consider factors such as offloading latency, energy consumption, reliability, and security to achieve a low-latency, low-energy, reliable, and secure task offloading strategy.

\section{System Model}\label{sec:System Model}
In this section, we first introduced the ISTN network model that we considered. Next, we presented a detailed task offloading framework, including the communication model, computational model, reliability model, and information security strength model. Finally, we formulated the problem. Table \ref{table1} summarizes the notations in our system.

\subsection{Integrated Satellite-Terrestrial Network}\label{sec:Integrated Satellite-Terrestrial Network}

\begin{figure}[!t]
\centering
\includegraphics[width=0.85\columnwidth]{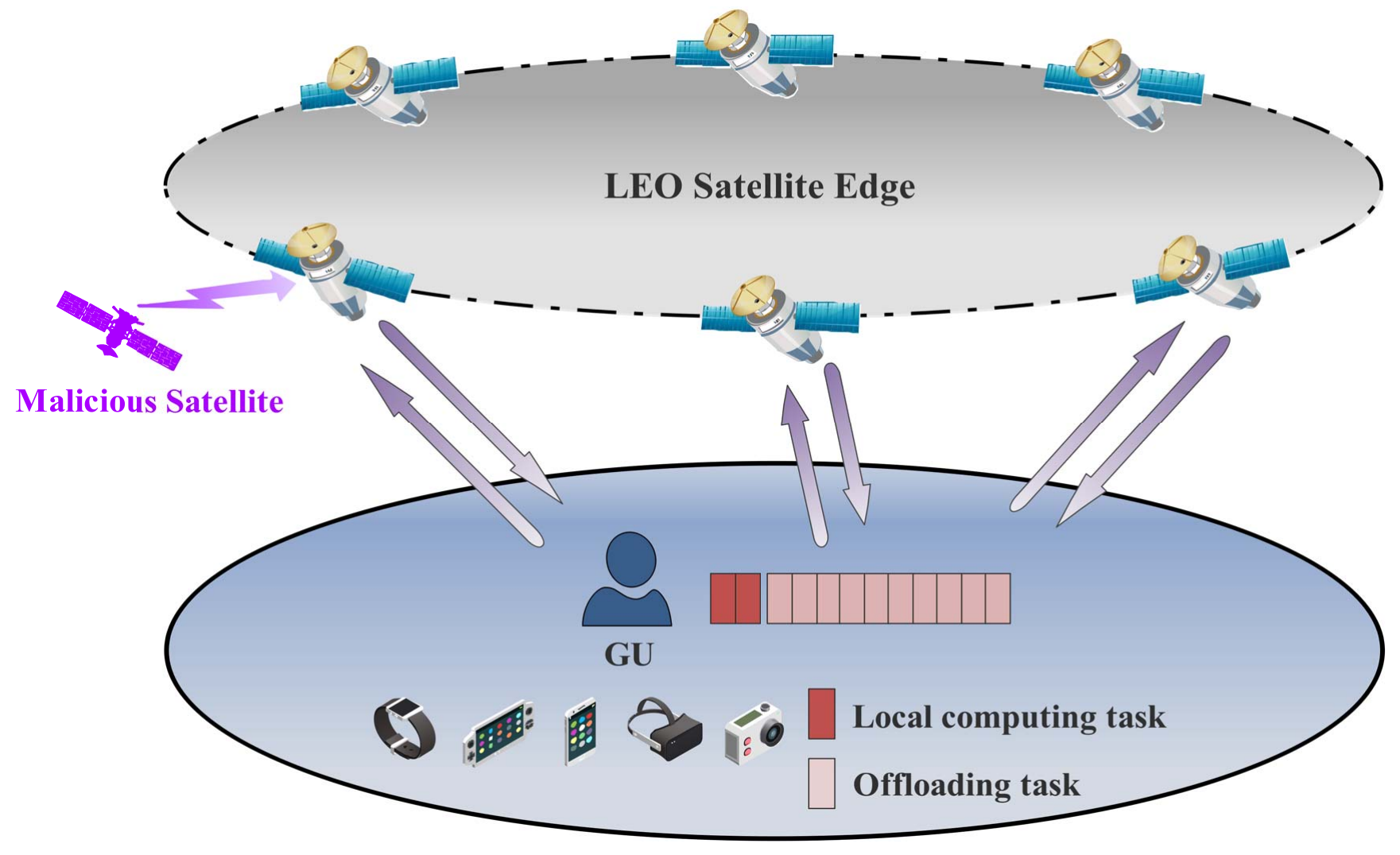}
\caption{Integrated satellite-terrestrial network.}
\label{fig:Integrated satellite-terrestrial network}
\end{figure}

We consider the deployment of an integrated satellite-terrestrial Network (ISTN) with LEO satellite edge, as shown in Figure \ref{fig:Integrated satellite-terrestrial network}. In this scenario, the ground user (GU) are limited in their computational capabilities and cannot process a large number of computing tasks on their own. To address this limitation, our system includes some LEO satellites moving uniformly on orbital planes. These LEO satellites are considered trusted entities and form the LEO satellite edge. Each LEO satellite is equipped with servers that can provide computational services for offloaded tasks. Thus, the LEO satellite edge can collaborate with GU to assist in task processing. In this situation, GU have the option to choose between two offloading schemes for their tasks: firstly, GU can choose to perform task computations on their local devices. Secondly, GU can choose to offload the tasks to the LEO satellite edge via a Ka-band radio link, and then complete the computation on the LEO satellite edge. Finally, the computational results will be sent back to the GU.

Besides, there exist malicious satellites within the communication range that possess the capability to attack the offloaded task data. Therefore, when the GU offloads computing tasks to the LEO satellite edge, there is a potential risk of the offloaded data being attacked. To effectively address this situation, we employ well-established and practical block ciphers to protect the information. Specifically, we divide each task into fixed-sized data blocks and subsequently apply independent encryption operations to each data block.

In the ISTN environment, time is discretized into $T$ scheduling periods, represented by the set $\mathcal{T}=\{1,\ldots,\tau,\ldots,T\}$. At the beginning of each scheduling period $\tau$, the GU generates $I$ security-sensitive computing tasks. The GU has the option to offload these computing tasks for process either on local devices or on the LEO satellite edge. Assuming that the LEO edge consists of $J$ LEO satellites, we represent these satellites with the set ${\mathcal{J}}=\{1,\ldots,j,\ldots,J\}$. Assuming that during the process of offloading tasks to the LEO satellite edge, the number of malicious satellites present within the communication range is $x$ and follows a Poisson distribution with parameter $\mu$, be represented as $x \sim \text{Poisson}(\mu)$. The set of $I$ security-sensitive computing tasks generated in each scheduling period $\tau$ is represented as ${\mathcal{I}}(\tau)=\{1,\ldots,i,\ldots, I\}$. The task $i$ generated in scheduling period $\tau$ is denoted as ${\mathcal{I}_{i}(\tau)} \stackrel{\Delta}{=} \{{D_{i}(\tau)},{S_{i}(\tau)}\}$, where $D_{i}(\tau)$ follows a Poisson distribution\cite{9237965} with parameter $\lambda$, denoted as $D_{i}(\tau) \sim \text{Poisson}(\lambda)$, and $\mathcal{S} _{i}(\tau)$ represents the security level requirement of the task. As each task may have different security requirements, we categorize these requirements into three levels: low, medium, and high levels.

In each scheduling period $\tau$, the GU's task allocation decision ${\eta_{ij}(\tau)} \in \{0,1\}$ represents whether the computing task $\mathcal{I}_{i}(\tau)$ is offloaded to the LEO satellite $j$. Since each computing task needs to be allocated only once, the GU's task allocation decision must satisfy the following constraints:

\begin{footnotesize}
\begin{equation} \sum _{j \in \mathcal {J}}  {\eta_{ij}(\tau)}  \le 1,\forall \tau \in \mathcal {T},\forall i \in \mathcal {I}.
\end{equation}
\end{footnotesize}

Particularly, when $\sum _{j \in \mathcal {J}} {\eta_{ij}(\tau)}=0$, it indicates that the computing task $\mathcal{I}_{i}(\tau)$ is processed on the local device.

The GU's local device and each LEO satellite respectively maintain a First-Come-First-Served (FCFS) queue to process the offloaded tasks. Additionally, due to the adoption of a sequential offloading scheme\cite{8876612, 8279411}, the order in which tasks are offloaded has a significant impact on the overall system performance. Therefore, besides making task allocation decisions, the GU also needs to determine the order in which tasks are offloaded. We represent the task offloading order as $\mathcal{O}(\tau)=\{o_{1}(\tau),o_{2}(\tau),{\ldots },o_{I}(\tau)\}$, where $o_{i}$ represents the offloading order of task $\mathcal{I}_{i}(\tau)$.

Therefore, in each scheduling period $\tau$, the task offloading decision made by the GU is represented as $A(\tau)=\{{\eta(\tau}),\mathcal{O}(\tau)\}$. It is important to note that the task allocation decision ${\eta(\tau)}$ can simultaneously represent which tasks to be offloaded and they are offloaded to which LEO satellite.

\begin{table}[!t] 
\centering
\caption{Summary of notations in our system.}
\footnotesize
\begin{tabular}{p{1.4cm}p{7cm}} 
\hline 
notation &  description\\
\hline
$ \mathcal{T}$ & {The set of scheduling periods} \\
$ T $ & {Total number of scheduling periods} \\
$ \tau $ & {Scheduling period} \\
$ \mathcal{I} $ & {Set of tasks} \\
$I$ & {Total number of tasks} \\
$i$ & {Task number} \\
$ {\mathcal{J}} $ & {Set of LEO satellites} \\
$ J $ & {Total number of LEO satellites} \\
$ j $ & {LEO satellites number} \\
$x$&{Number of malicious satellites}\\
$ \mu $ & {Average number of malicious satellites}\\
$ D_{i} $ & {Data size of task $\mathcal{I}_{i}$} \\
$ \mathcal{S} _{i} $ & {Security level requirement of the task $\mathcal{I}_{i}$} \\
$\mu$ & {Average number of data sizes} \\
$ A(\tau) $ & {Task offloading decision in scheduling period $\tau$} \\
$ \eta(\tau) $ & {Task allocation decision in scheduling period $\tau$} \\
$ \mathcal{O}(\tau) $ & {Task offloading order in scheduling period $\tau$} \\

$ s_{j} $ & {Distance between the GU and the LEO satellite $j$} \\
$ h_{j}^{Ka} $ & {Channel gain received by the LEO satellite $j$} \\
$ SNR_{j} $ & {SNR of the link between the GU and satellite $j$} \\
$ P^{Ka} $ & {Transmission power of the GU} \\
$ \hat{\sigma}^2 $ & {Variance of the AWGN} \\
$ R_{j}^{LEO} $ & {Achievable transmission rate between the GU and the LEO satellite $j$} \\
$ BER_{j} $ & {BER of the transmission from the GU to satellite $j$} \\

$ q^{local} $ & {Average number of CPU cycles required for the GU to process 1 bit} \\
$ f^{local} $ & {Local computation frequency of the GU} \\
$ T_{i}^{local }$ & {Total latency for the GU to perform local computation for task $\mathcal{I}_{i}$} \\
$ T^{wait}_{i} $ & {Waiting time of task $\mathcal{I}_{i}$} \\
$ t^{local,start}_{i} $ & {The time of GU starts to process task $\mathcal{I}_{i}$ on the local device} \\
$ t^{local,end}_{i} $ & {Completion time of task $\mathcal{I}_{i}$ on the local device } \\
$ q^{en} $ & {Average number of CPU cycles for the GU to encrypt 1 bit} \\
$ f^{en} $ & {Encryption computation frequency of the GU} \\
$ T^{en}_{i} $ & {Encryption time for the GU to encrypt task $\mathcal{I}_i$} \\
$ T_{ij}^{tran} $ & {Transmission time to offload task $\mathcal{I}_i$ to LEO satellite $j$} \\

$ T_{ij}^{comp} $ & {Computation time for the LEO satellite $j$ to process task $\mathcal{I}_i$} \\
$ T_{ij}^{LEO} $ & {Total latency for computation task $\mathcal{I}_{i}$ to be offloaded and processed by the LEO satellite $j$} \\
$T_{ij}^{wait}$& {Waiting time of task $\mathcal{I}_{i}$ on the LEO satellite $j$} \\
$ t^{LEO,start}_{ij} $ & {The time of GU starts to offloading task $\mathcal{I}_{i}$} \\
$ t^{LEO,end}_{ij} $ & {Completion time of task $\mathcal{I}_{i}$ on the LEO satellite $j$ } \\
$ E^{local}_{i} $ & {Computational energy consumption  for the GU to process local task $\mathcal{I}_{i}$} \\

$ E_{total}^{local} $ & {Total energy consumption for the GU to process all local tasks} \\
$ k $ & {Coefficient representing the hardware architecture of the local device} \\
$ E_{ij}^{LEO }$ & {Energy consumption for processing the computation task $\mathcal{I}_{i}$ at the LEO satellite edge} \\
$ E_{total}^{LEO }$ & {Total energy consumption for processing all tasks at the LEO satellite edge} \\
$ r_{ij} $ & {Probability of successful offloading for task $\mathcal{I}_{i}$} \\
$ r_{total} $ & {Probability of GU offloading successfully} \\
$ N $ & {Block cipher length} \\
$ \Phi_{i} $ & {Successful probability of malicious satellites' attacks on a block cipher with a length of $N_i$} \\

$ S_i $ & { Information security strength of task $\mathcal{I}_i$} \\
$ A_{i}  $ & {Situation of task $\mathcal{I}_{i}$ being attacked} \\
$ \xi $ & {Random number} \\
$ T_{total} $ & {Total offloading delay} \\
$ E_{total} $ & {Total energy consumption} \\
$ A_{total} $ & {Total number of attacks} \\
$ \rho $ & {Probability of successful task offloading} \\
$ \beta_{1},\beta_{2} $ & {Weight factors} \\

\hline
\end{tabular}
\label{table1}    
\end{table}

\subsection{LEO Satellite Communication}\label{sec:LEO satellite communication}
\begin{figure}[!t]
	\centering
	\includegraphics[width=0.85\columnwidth]{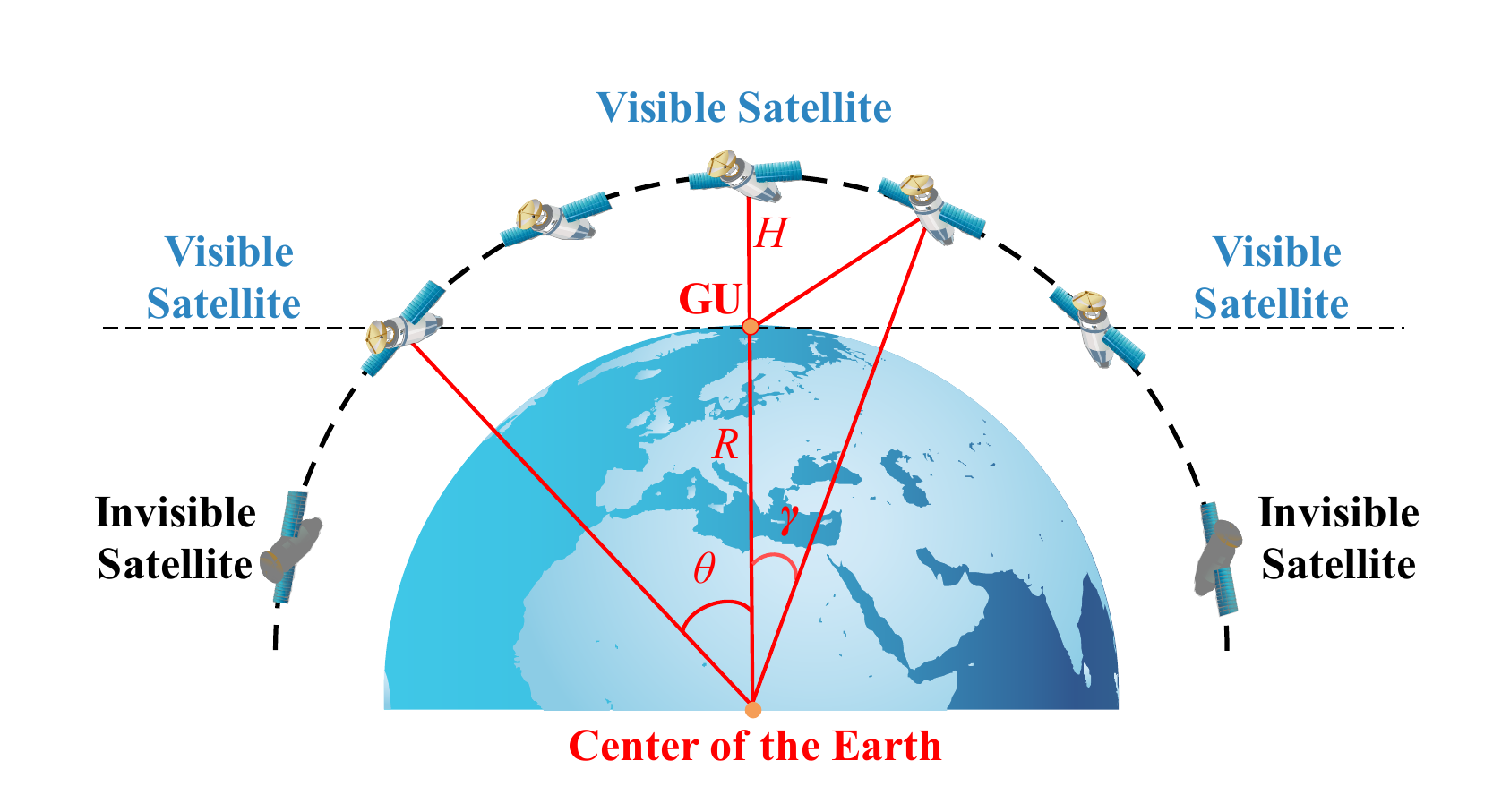}
	\caption{Satellite coverage model.}
	\label{fig:Satellite Coverage Model}
\end{figure}

For LEO satellite communication, the GU can obtain information about the altitude, velocity, and position of the LEO satellite within a scheduling period since the orbit is pre-planned. Due to the high mobility of LEO satellites, the channel conditions are constantly changing. Additionally, each LEO satellite can only provide service to the GU within a certain range of positions \cite{9344666, sym14030564}. Since line-of-sight (LOS) transmission is primarily considered, the spatial relationship between the LEO satellite edge and the GU is illustrated in Figure \ref{fig:Satellite Coverage Model}.

In LOS transmission, we assume that the angle between the line connecting satellite $j$ and the center of the Earth and the reference line is denoted as $\gamma$. The boundary angle for the visible range is represented as $\theta$. Therefore, task offloading to the LEO satellite $j$ is only possible when the absolute value of $\gamma$ is less than the boundary value $\theta$. Otherwise, the task offloading will fail.

When a computing task is offloaded to the LEO satellite $j$, according to the cosine rule, the distance between the GU and the LEO satellite $j$ can be calculated as 
$
s_{j}=\sqrt{R^2+(R+H)^2-2R(R+H)\cos(\gamma)},
$
where $R$ is the radius of the Earth and $H$ is the orbital height of the LEO satellite edge.

To simplify the analysis, we consider the free-space path loss (FSPL) model \cite{9583851, 8611345} and additive white Gaussian noise (AWGN). Therefore, the channel gain received by the LEO satellite $j$ in the Ka-band can be represented as 
$
h_{j}^{Ka}= \frac {\beta _{o}}{{s_{j}}^{2}},
$
where $\beta _{o}$ is the parameter for the channel gain.

To ensure the communication quality and reliability between LEO satellites, it is assumed that each satellite uses an independent frequency band during transmission to avoid signal interference. The signal-to-noise ratio (SNR) of the link between the GU and satellite $j$ can be represented as 
$
	SNR_{j}=\frac {P^{Ka}{{h_{j}^{Ka}}}} {\hat {\sigma }^{2}},
$
where $P^{Ka}$ is the transmission power of the GU and $\hat{\sigma}^2$ is the variance of the AWGN.

According to Shannon's formula, the achievable transmission rate between the GU and the LEO satellite $j$ in the Ka-band can be represented as 
$
	R_{j}^{LEO}=  {B_{Ka}}\log _{2}\left ({1+\frac {P^{Ka}{h_{j}^{Ka}}} {\hat {\sigma }^{2}} }\right),
$
where $B_{Ka}$ is the total bandwidth in the Ka-band.

In our research, we use the bit error rate (BER) as a metric to measure the probability of errors in the transmission of 1 bit of data. Assuming that the satellite communication system adopts binary phase-shift keying (BPSK) modulation, the BER of the transmission from the GU to satellite $j$ can be calculated as 
$
BER_{j}=\frac{erfc(\sqrt{SNR_{j}} )}{2},
$
where $erfc()$ represents the complementary error function.

\subsection{Computational Model}

Our system integrates local computation and LEO satellite edge computation, providing the GU with two different computation offloading schemes that differ in terms of offloading time and energy consumption. In this section, we will focus on discussing the computation tasks generated by the GU in a specific scheduling period $\tau$. To simplify the representation, we will ignore specific descriptions and relevant symbols related to the scheduling period $\tau$.

\subsubsection{Local Computation}

When the GU performs local computation for task $\mathcal{I}_{i}$, there is no need for additional encryption time as block cipher protection is not required. The average number of CPU cycles required for the GU to process 1 bit is denoted as $q^{local}$, and the local computation frequency of the GU is represented as $f^{local}$.

The GU's local device maintains an FCFS service queue and processes local tasks in order. If the GU's local device's waiting queue is not empty, the task needs to enter the waiting queue and wait for the previous tasks to be processed before it can start processing. Therefore, the total latency required for the GU to perform local computation for task $\mathcal{I}_{i}$ can be represented as:

\begin{footnotesize}
	\begin{equation}
T_{i}^{local} =
\begin{cases}
	T^{wait}_{i}+\frac{D _{i}q^{local}}{f^{local}}, & \text{if } \eta_{ij}\neq 1 ,\forall  j \in \mathcal{J}, \\
	0, & \text{otherwise},
\end{cases}
	\end{equation}
\end{footnotesize}
where $T^{wait}_{i}$ represents the waiting time of task $\mathcal{I}_{i}$ on the local device.

Assuming that the GU starts to process local computation task $\mathcal{I}_{i}$ at time $t^{local,start}_{i}$, the completion time of task $\mathcal{I}_{i}$ on the local device can be represented as $t^{local,end}_{i}=t^{local,start}_{i}+T_{i}^{local}$.

\subsubsection{LEO Satellite Edge Computation}
When the GU's task $\mathcal{I}_i$ is offloaded to the LEO satellite edge, we need to consider the time for encryption, transmission, and computation at the LEO satellite edge.

Before offloading, task $\mathcal{I}_i$ needs to be encrypted to ensure a certain level of transmission security. The average number of CPU cycles required for the GU to encrypt 1 bit is denoted as $q^{en}$, and the encryption computation frequency of the GU is represented as $f^{en}$. Therefore, the encryption time required for the GU to encrypt task $\mathcal{I}_i$ is calculated as $T^{en}_{i}=\frac{D_{i}q^{en}}{f^{en}}$. If task $\mathcal{I}_i$ is offloaded to the LEO satellite $j$, the use of block ciphers does not increase the size of the transmitted data. Therefore, the transmission time can be expressed as $T_{ij}^{tran} = \frac{D_i}{R_j^{LEO}}.$

\begin{figure}[!t]
	\centering
	\includegraphics[width=0.85\columnwidth]{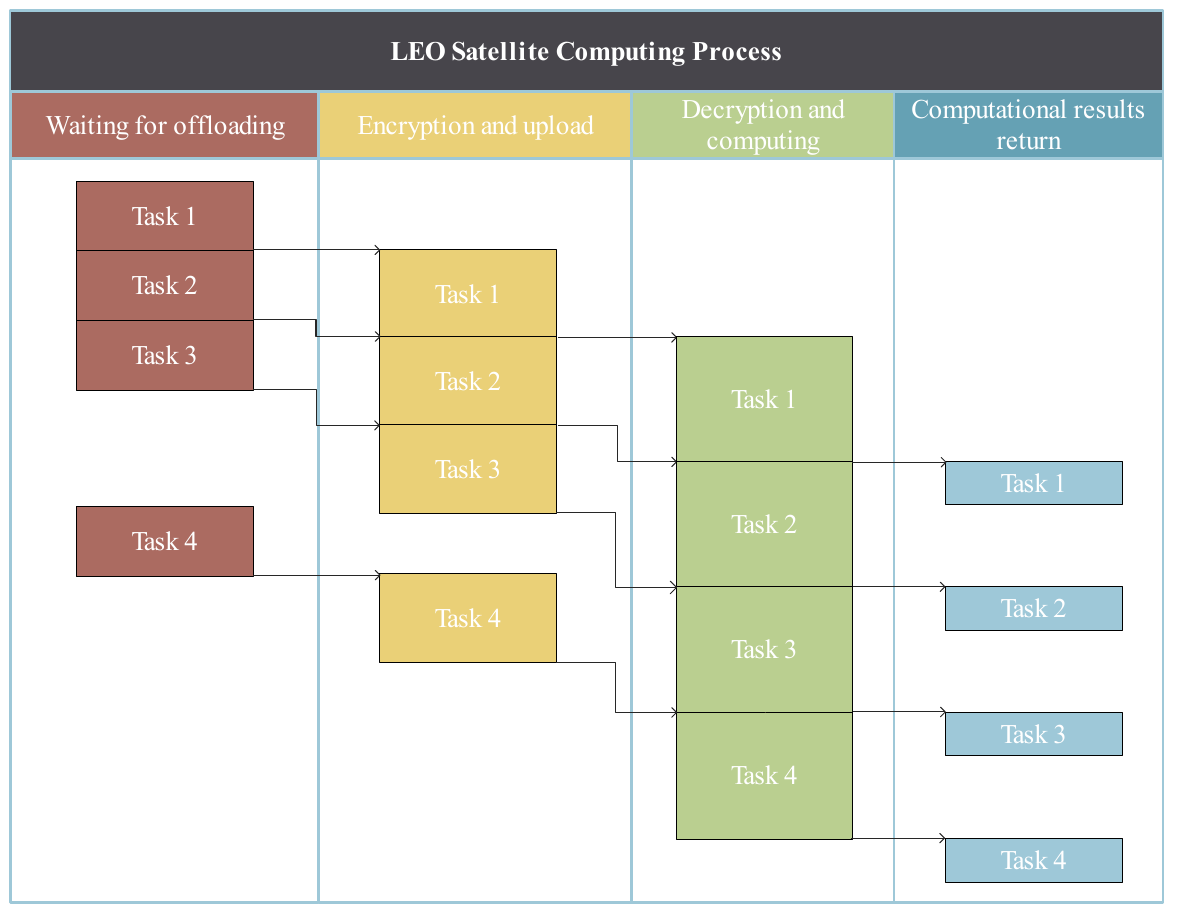}
	\caption{LEO satellite computing process.}
	\label{fig:LEO satellite computing process}
\end{figure}

Each LEO satellite uses a separate FCFS queue to process offloaded tasks. Therefore, after task $\mathcal{I}_{i}$ is transmitted to the LEO satellite $j$, if the server's waiting queue of the LEO satellite $j$ is not empty, task $\mathcal{I}_{i}$ needs to enter the waiting queue and wait for the previous tasks to be processed before it can start processing. The specific process is shown in Figure \ref{fig:LEO satellite computing process}.

The average number of CPU cycles required for the LEO satellite $j$ to process 1 bit is denoted as $q_{j}$, and the computation frequency of the LEO satellite $j$ at the LEO edge is represented as $f_{j}^{LEO}$. Therefore, the computation time required for the LEO satellite $j$ to process task $\mathcal{I}_i$ is calculated as $T_{ij}^{comp}=\frac {D_{i}q_{j}}{f_{j}^{LEO}}$.

Based on the above discussion, the total latency $T_{ij}^{LEO}$ required for computation task $\mathcal{I}_{i}$ to be offloaded and processed by the LEO satellite $j$ can be represented as:

\begin{footnotesize}
	\begin{equation}
T_{ij}^{LEO} =
\begin{cases}
	\frac{D_{i}q^{en}}{f^{en}} + \frac {D_{i}} {R_{j}^{LEO}} +T_{ij}^{wait}+\frac {D_{i}q_{j}}{f_{j}^{LEO}}, & \text{if } \eta_{ij}= 1 ,\exists j \in \mathcal{J}, \\
	0, & \text{otherwise},
\end{cases}
	\end{equation}
\end{footnotesize}
where $T_{ij}^{wait}$ represents the waiting time of task $\mathcal{I}_{i}$ on the LEO satellite $j$. This waiting time depends on the arrival time of task $\mathcal{I}_{i}$ and the waiting queue on the LEO satellite $j$, which varies dynamically with the system operation. Compared to the encryption and uplink transmission latency of the GU, the decryption and downlink transmission latency of the satellite server can be ignored.

The completion time of task $\mathcal{I}_{i}$ on the LEO satellite $j$, generated in a scheduling period $\tau$, depends on the completion time of the tasks already received on the LEO satellite $j$ and their respective computation times. Assuming that the GU starts to offloading task $\mathcal{I}_{i}$ at time $t^{LEO,start}_{ij}$, the completion time of task $\mathcal{I}_{i}$ on the LEO satellite $j$ can be represented as $t^{LEO,end}_{ij}=t^{LEO,start}_{ij}+T_{ij}^{LEO}$.

\subsection{Energy Consumption Model}

\subsubsection{Local Energy Consumption}
The computational energy consumption required for the GU to process local task $\mathcal{I}_{i}$ can be represented as:

\begin{footnotesize}
	\begin{equation}
E^{local}_{i}=k{f^{local}}^{3}t^{local}_{i}, \nexists \eta_{ij}=1,\forall i \in {\mathcal{ I}}, \forall j \in {\mathcal{ J}},
	\end{equation}
\end{footnotesize}
where $k$ is a coefficient representing the hardware architecture of the local device.

The total energy consumption required for the GU to process all local tasks can be represented as:

\begin{footnotesize}
	\begin{equation}
E_{total}^{local} = \sum _{i \in \mathcal {I}} \sum _{j \in \mathcal {J}} E_{ij}^{local}.
	\end{equation}
\end{footnotesize}

\subsubsection{LEO Satellite Edge Energy Consumption}
When the GU offloads computation task $\mathcal{I}_{i}$, the energy consumption required for encryption is $k(f^{en})^3T^{en}_{i}$, and the energy consumption for transmission is $P^{Ka}\frac {D_{i}} {R_{j}^{LEO}}$. Therefore, the energy consumption for processing the computation task $\mathcal{I}_{i}$ at the LEO satellite edge can be represented as:

\begin{footnotesize}
	\begin{equation}
E_{ij}^{LEO} =  k{f^{en}}^{3}T^{en}_{i}+P^{Ka}\frac {D_{i}} {R_{j}^{LEO}}, \exists \eta_{ij}=1,\forall i \in {\mathcal{ I}}, \forall j \in {\mathcal{ J}}.
\end{equation}
\end{footnotesize}

The total energy consumption required for processing all tasks at the LEO satellite edge can be represented as:

\begin{footnotesize}
	\begin{equation}
E_{total}^{LEO}=\sum _{i \in \mathcal {I}} \sum _{j \in \mathcal {J}}E_{ij}^{LEO}.
\end{equation}
\end{footnotesize}

\subsection{Reliability Model}
In the satellite communication environment, the high mobility of satellites can result in adverse wireless transmission conditions for the \textit{satellite-terrestrial} link, leading to potential failures in task offloading. Therefore, when devising offloading strategies, it is crucial to consider the reliability of the transmission, which is defined as the probability of offloading failure\cite{8279411}. During the task offloading process, it is necessary to ensure that the probability of offloading failure is controlled within a certain range to guarantee the reliability of the transmission. Assuming that the GU offloads computation task $\mathcal{I}_{i}$ to the LEO satellite $j$, the BER during the wireless communication transmission process is denoted as $BER_{ij}$. For each task, every 1 bit of data must be transmitted correctly to consider the task offloading successful.

Furthermore, as described in Section \ref{sec:LEO satellite communication}, each LEO satellite can only provide services to the GU within a certain coverage area. Therefore, when the GU chooses to offload task $\mathcal{I}_{i}$ to the LEO satellite $j$, if the absolute value of $\gamma$ is greater than or equal to the boundary value $\theta$, it will also result in offloading failure. Once task $\mathcal{I}_i$ computation is completed, if the LEO satellite $j$ moves out of the LOS, the LEO edge can migrate the computation result to another LEO satellite within the LOS using inter-satellite links (ISL). Then, the computation result can be sent back, ensuring the successful transmission of the computation result. Therefore, the probability of successful offloading for task $\mathcal{I}_{i}$ can be represented as:

\begin{footnotesize}
	\begin{equation}
r_{ij} = \begin{cases} (1-BER_{ij}) ^{D_{i}}, \exists \eta_{ij}=1,\mid \gamma \mid < \theta,\forall i \in {\mathcal{ I}}, \forall j \in {\mathcal{ J}}, \\ 1, \text {otherwise}. \end{cases}
\end{equation}
\end{footnotesize}

For the $I$ computation tasks in the scheduling period $\tau$, it is considered a successful offloading for the GU only if every task is transmitted correctly. Otherwise, it is considered as a failure offloading. It is important to note that local computation tasks can always be offloaded successfully since they do not require wireless transmission.

In summary, the probability of GU offloading successfully can be represented as: 

\begin{footnotesize}
	\begin{equation}
r_{total}=\prod _{i \in \mathcal {I}} \prod _{j \in \mathcal {J}} {r_{ij}}.
	\end{equation}
\end{footnotesize}

\subsection{Information Security Strength Based On Adversary Model}
In general, the security level can be measured by factors such as the mathematical complexity of the encryption scheme and the length of the encryption key. However, different encryption algorithms may be vulnerable to various attack methods, making it difficult to accurately measure security using universal standards.

When evaluating the security of a system, we can measure the security level by assessing the adversary's capability to break block ciphers. Specifically, the ability of an attacker to break a block cipher of a specific block length is related to the probability mass function (PMF). We introduce the parameter \textit{attacker strength} denoted as $\sigma$, which is associated with the length of the encryption block. The probability that an attacker can break a block cipher of length $N$ can be represented as $P_{r}(\sigma = N)$. An attacker with strength $\sigma$ is capable of breaking any block cipher with a length less than or equal to $\sigma$. Therefore, $P_{r}(\sigma = N)$ can also be viewed as the probability of the block cipher being broken by an attacker, leading to data leakage. It can be expressed as $\Phi=P_{r}(\sigma=N)=P_{r}(\sigma \geq N)$.

When the GU offloads tasks to the LEO edge, the risk of block cipher being attacked needs to be considered if there are malicious satellites within the communication range. In this scenario, to ensure the security of the system, we have adopted the RIJNDAEL algorithm as the encryption algorithm, which is a popular symmetric encryption algorithm. We have defined the maximum block length $N_{max}$ and the minimum block length $N_{min}$ available in the cryptographic system. Due to the use of the RIJNDAEL encryption algorithm, the block length is restricted to between 128 and 256 bits, and it is a multiple of 32 bits. Therefore, $N_{max}$ is set to 256, and $N_{min}$ is set to 128. Based on the security level of the tasks, we have determined the encryption block lengths for different security-level tasks, as follows: For tasks with a low-security level, the encryption block length is set to 192 bits. For tasks with a medium-security level, the encryption block length is set to 224 bits. For tasks with a high-security level, the encryption block length is set to 256 bits. 

We employed a linear adversary strength model to represent the attack intensity of malicious satellites\cite{10102414, 4358705}, which can be expressed as $\frac{1}{N_{max }-N_{min }}.$

We assume that the successful probability of malicious satellites' attacks on a block cipher with a length of $N_i$ follows a uniform distribution, represented as $\Phi_{i}=P_{r}\left(\sigma \geq N_{i}\right)=\frac{N_{max }-N_{i}}{N_{max }-N_{min }}.$

To measure the information security strength of a task, we consider the presence of malicious satellites within the communication range. Let's assume that during the process of task $\mathcal{I}_{i}$ offloading to satellite $j$, there are $x$ malicious satellites within the communication range. The information security strength of task $\mathcal{I}_i$ can be defined as the probability of $x$ malicious satellite attacks failing, which represents the probability of task $\mathcal{I}_i$ successfully avoiding attacks, which can be expressed as $S_i=(1-\Phi_{i})^{x},x \sim Poisson(\mu).$

Whether task $\mathcal{I}_i$ is attacked successfully (i.e., does not satisfy $S_i$) is an independent random event. To estimate the total number of successful attacks on tasks, denoted as $A$, we employ a Monte Carlo simulation\cite{9651535, 9237965}. For task $\mathcal{I}_{i}$, the determination of whether it is successfully attacked is based on its information security strength $S_i$. If it is successfully attacked, the number of attacks is increased by 1. Thus, the situation of task $\mathcal{I}_{i}$ being attacked can be expressed as:

\begin{footnotesize}
\begin{equation}
A_{i} =
\begin{cases}
	1, & \text{if } \xi > S_i , \\
	0, & \text{otherwise},
\end{cases}
\end{equation}
\end{footnotesize}
where $\xi$ is a random number between 0 and 1.

\subsection{Problem Formulation}

Since local computation and LEO satellite edge computation progress simultaneously, the overall offloading delay depends on the larger value between the completion time of local computation and the completion time of LEO satellite edge computation. The total offloading delay can be expressed as:

\begin{footnotesize}
\begin{equation}
T_{total} = \max \left( \max_{\tau \in \mathcal{T},{i \in \mathcal{I}}} T^{local,end}_{i}(\tau), \max_{\tau \in \mathcal{T},{i \in \mathcal{I}},{j \in \mathcal{J}}}  T^{LEO,end}_{ij}(\tau) \right).
\end{equation}
\end{footnotesize}
In this equation, the first term in the $\max()$ function represents the latest completion time among all local computation tasks. The second term represents the latest completion time among all offloading tasks in all scheduling periods.

The total energy consumption can be expressed as:

\begin{footnotesize}
\begin{equation}
\begin{aligned} E_{total}=&E_{total}^{local} + E_{total}^{LEO} \\=&\sum _{{\tau \in \mathcal {T}},i \in \mathcal {I},{j \in \mathcal {J}}}\left ({E_{ij}^{local}(\tau)+E_{ij}^{LEO}(\tau)}\right).\end{aligned}
\end{equation}
\end{footnotesize}

By accumulating the number of times each task is attacked in each scheduling period, we can obtain an estimate for the total number of attacks, denoted as $A_{total}$, which can be expressed as:

\begin{footnotesize}
\begin{equation}
A_{total}=\sum _{{\tau \in \mathcal {T}},i \in \mathcal {I}}A_{i}(\tau).
\end{equation}
\end{footnotesize}

Based on the aforementioned definitions, the optimization objective can be formulated as the minimization of a weighted sum of the offloading delay, energy consumption, and the number of attacks, which is referred to as the system cost and denoted as:

\begin{footnotesize}
\begin{equation}
\begin{aligned} \mathbf{SP1}: \underset {\{\eta ,  \mathcal{O}\}}{\mathrm{ min}}& (T_{total}+\beta_{1}E_{total}+\beta_{2}A_{total}) \\{\mathrm{s}}{\mathrm{.t}}{\mathrm{.}} \qquad  &{\eta_{ij}(\tau)} \in \{ {0,1}\},\forall \tau \in {\mathcal{ T}}, \forall i \in {\mathcal{ I}}, \forall j \in {\mathcal{ J}}, \\ &\sum _{j \in \mathcal {J}}  {\eta_{ij}(\tau)}  \le 1, \forall \tau \in \mathcal {T},\forall i \in \mathcal {I},   
\\  &r_{total}(\tau)\ge{\rho },\forall \tau \in \mathcal {T},\end{aligned}
\end{equation}
\end{footnotesize}
where $\beta_{1}$ and $\beta_{2}$ are weight factors used to balance the importance of offloading delay, energy consumption, and the number of attacks. Under this optimization objective, we need to consider several constraint conditions. The first and second constraint conditions ensure that each task in each scheduling period can only select either a LEO satellite or local device for offloading. The third constraint ensures transmission reliability, requiring the probability of successful task offloading $r_{total}(\tau)$ in each scheduling period to be greater than or equal to a threshold value $\rho$.

\section{Security-Sensitive Task Offloading Solution}\label{sec:Security-Sensitive Task Offloading Solution}
In this section, we have established a Markov decision process (MDP) model for security-sensitive task offloading in the ISTN environment. Then, we introduced a security-sensitive task offloading strategy optimization algorithm based on PPO for solving this model.

\subsection{Markov Decision Process Model}
In the ISTN environment, the task offloading problem exhibits Markovian properties, allowing us to model the task offloading problem as an MDP. In the MDP model for task offloading in the ISTN environment, the agent is the GU, and the environment consists of elements related to task offloading in the ISTN. These elements include generated task information and LEO satellite edge status information. Time in the ISTN environment is discretized into several scheduling periods. At the beginning of each scheduling period, the GU obtains the state for that period from the environment. Then, based on the offloading strategy, it makes task offloading decisions. After executing these decisions, the ISTN environment changes, and new environmental information serves as reward feedback for the GU to evaluate the quality of the offloading decisions. Following the changes in the ISTN environment, the system enters the next scheduling period. The GU repeats these actions until all scheduling periods have concluded. The state space, action space, and immediate reward in the ISTN environment are described as follows:

\textbf{State space:} The state of the ISTN environment serves as the foundation for the GU to make task offloading decisions. The GU must base its offloading decisions on the current scheduling period $\tau$, including the task information, LEO satellite edge load status, and the current time. The ISTN environment's state during scheduling period $\tau$ is defined as $S(\tau) = \{{\mathcal{I(\tau)}}, L(\tau), t\}$, where $\mathcal{I(\tau)}$ represents the information of the $I$ security-sensitive computing tasks generated in scheduling period $\tau$, $L(\tau)$ corresponds to the load status information of LEO satellite edges, and $t$ denotes the moment when scheduling period $\tau$ begin.
		
\textbf{Action space:} The action space defines the task offloading decisions made by the GU during the scheduling period $\tau$. This includes which tasks should be offloaded, to which specific LEO satellite they should be offloaded, and their offloading order. Hence, the GU's offloading actions are defined as $A(\tau) = \{{\eta(\tau}), \mathcal{O}(\tau)\}$, where, as defined in Section \ref{sec:Integrated Satellite-Terrestrial Network}, $\eta(\tau)$ denotes the task allocation decision, specifying which tasks should be offloaded and to which LEO satellite, while $\mathcal{O}(\tau)$ represents the offloading order of tasks.

\textbf{Immediate Reward:} The immediate reward is the feedback from the ISTN environment regarding the task offloading decision made by the GU. Since the optimization objective $\mathbf{SP1}$ is a minimization problem, we can use the negation of this optimization objective as the immediate reward. Therefore, the immediate reward for scheduling period $\tau$ can be defined as:
$r(\tau) = -(T_{total}(\tau) + \beta_{1}E_{total}(\tau) + \beta_{2}A_{total}(\tau)).$

\subsection{PPO-based Security-Sensitive Task Offloading Algorithm}

Reinforcement learning is an effective machine learning algorithm that seeks to maximize rewards through continuous interaction between an agent and its environment, making it well-suited for solving Markov decision problems. Furthermore, deep reinforcement learning combines deep learning techniques, such as deep neural networks (DNNs), with reinforcement learning, enhancing the efficiency and reliability of the decision-making process. Schulman et al. \cite{schulman2017proximal} introduced the proximal policy optimization (PPO) algorithm, which is a state-of-the-art deep reinforcement learning method applicable to both discrete and continuous state and action spaces. Moreover, it outperforms other algorithms on benchmark tests while offering a good balance between ease of adjustment, efficient sampling, and straightforward implementation. Therefore, we choose the PPO algorithm to generate task offloading strategies and propose a security-sensitive task offloading strategy optimization algorithm based on PPO.

\begin{figure}[!t]
\centering
\includegraphics[width=0.85\columnwidth]{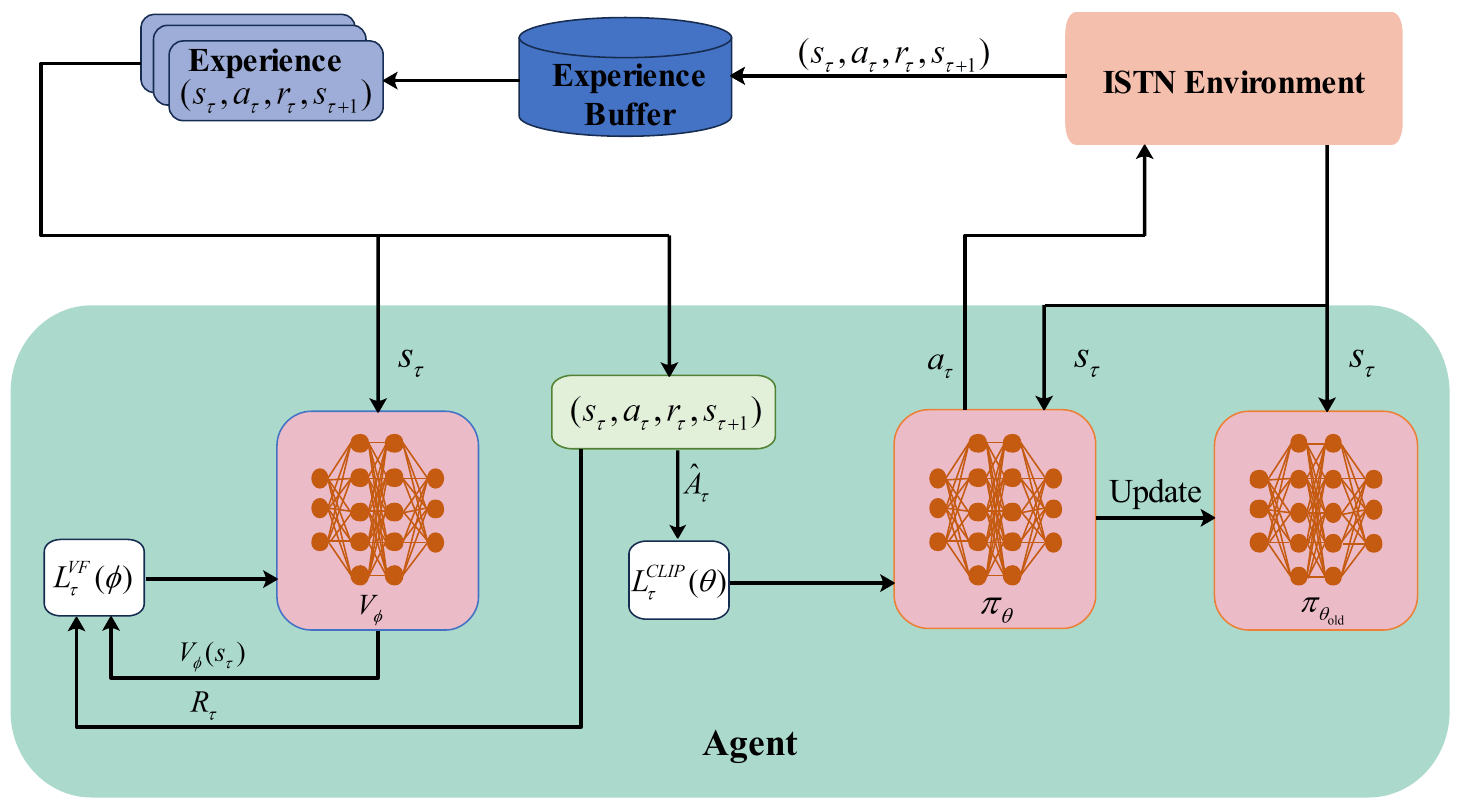}
\caption{The PPO algorithm framework.}
\label{fig:The PPO algorithm framework}
\end{figure}

PPO is a policy gradient algorithm that optimizes task offloading strategies by optimizing the policy network. The PPO-based deep reinforcement learning (DRL) training framework is illustrated in Figure \ref{fig:The PPO algorithm framework}. Specifically, it consists of three neural network components: the new policy network $\pi_{\theta}$, the old policy network $\pi_{\theta_{\text {old }}}$, and the value network $V_{\phi}$. PPO interacts with the environment using the old policy network $\pi_{\theta_{\text {old }}}$. In each scheduling period $\tau$, based on the current state $s_\tau$, an action $a_\tau$ is sampled from the old policy network $\pi_{\theta_{\text {old }}}$ and executed. The resulting next state $s_{\tau+1}$, immediate reward $r_\tau$, and other environmental information are observed. Finally, a batch of trajectories in the form $(s_\tau, a_\tau, r_\tau,s_{\tau+1})$ is generated and stored in the experience buffer $\mathcal{D}$.

PPO employs two loss functions to optimize the policy network and the value network, namely the policy loss function $L^{CLIP}_{\tau}(\theta)$ and the value loss function $L^{VF}_{\tau}(\phi)$ :

\textbf{Policy loss function :} The policy loss function $L^{CLIP}_{\tau}(\theta)$ measures the improvement of the new policy relative to the old policy and is expressed as $L^{CLIP}_{\tau}(\theta)=\mathbb{E}_{\tau}\left[\min \left(\frac{\pi_{\theta}\left(a_{\tau} \mid s_{\tau}\right)}{\pi_{\theta_{\text {old }}}\left(a_{\tau} \mid s_{\tau}\right)} \hat A_{\tau}, \operatorname{clip}\left(\frac{\pi_{\theta}\left(a_{\tau} \mid s_{\tau}\right)}{\pi_{\theta_{\text {old }}}\left(a_{\tau} \mid s_{\tau}\right)}, 1-\epsilon, 1+\epsilon\right) \hat A_{\tau}\right)\right]$, where $\theta$ represents the parameters of the new policy network, $\pi_{\theta}(a_{\tau}|s_{\tau})$ is the probability of selecting action $a_{\tau}$ under the new policy, $\pi_{\theta_{\text{old}}}(a_{\tau}|s_{\tau})$ is the probability of selecting action $a_{\tau}$ under the old policy, $\hat A_{\tau}$ is the advantage function, and $\epsilon$ is a constant that controls the magnitude of policy updates.

\textbf{Value loss function :} The value loss function $L^{VF}_{\tau}(\phi)$ is used to evaluate the performance of the value network $V{\phi}$. The objective is to minimize the discrepancy between the estimated value by the value network and the actual return $R_{\tau}$, and it is expressed as  $L^{VF}_{\tau}(\phi)=\mathbb{E}_{\tau}\left[\left(V_{\phi}\left(s_{\tau}\right)-R_{\tau}\right)^{2}\right]$, where $\phi$ represents the parameters of the value network, $V_{\phi}(s_{\tau})$ is the estimated value of state $s_{\tau}$ by the value network, and $R_{\tau}$ is the observed reward from the experience.

The total loss function $L_{\tau}(\theta, \phi)$ of PPO is a linear combination of the policy loss and the value loss, and is expressed as $L_{\tau}(\theta, \phi) = L^{CLIP}_{\tau}(\theta) - c_1 L^{VF}_{\tau}(\phi) + c_2 S_{\pi_\theta}(s_\tau)$, where $S_{\pi_\theta}(s_\tau)$ denotes the entropy reward, which is used to promote exploration in the policy. Hyperparameters $c_1$ and $c_2$ are used to balance the importance of these three components.

The PPO algorithm continuously interacts with the environment, collects generated trajectories, and stores them in the experience buffer $\mathcal{D}$. Periodically, data is sampled from the experience buffer $\mathcal{D}$ and the parameters of the policy network $\pi_{\theta}$ and the value network $V_{\phi}$ are updated using gradient descent to minimize the total loss function $L(\theta,\phi)$. It is important to note that the training process of PPO usually involves multiple iterations, where the data in the experience buffer $\mathcal{D}$ is cleared in each iteration to obtain a better policy. By continuously optimizing the parameters $\theta$ of the policy network and $\phi$ of the value network, the optimization objective of secure-sensitive task offloading strategies can be achieved.

The pseudo-code for the training process of the PPO-based algorithm is presented in Algorithm \ref{alg1}.

\begin{algorithm}[!t]
\caption{Training process of the PPO-based algorithm.}
\footnotesize
\label{alg1}
\textbf{Input:} Environment dynamics, neural network architectures, hyperparameters \\
\textbf{Output:} Optimal policy $\pi^*$
\begin{algorithmic}[1]
\STATE Initialize policy network $\pi_{\theta}(a|s)$ and value network $V_{\phi}(s)$ with random weights\;
\STATE Initialize a replay buffer $\mathcal{D}$ for experience storage\;
\STATE Initialize hyperparameters $\epsilon$, $\beta$, and others\;

\FOR{$n = 1$ to $N_{\text{episodes}}$}
\STATE Initialize environment state $s_0$
\FOR{$\tau = 1$ to $T$}
\STATE Sample action $a_\tau \sim \pi_{\theta}(a|s_\tau)$
\STATE Execute action $a_\tau$, observe reward $r_\tau$ and next state $s_{\tau+1}$
\STATE Store $(s_\tau, a_\tau, r_\tau,s_{\tau+1})$ as a trajectory
\STATE $s_\tau \leftarrow s_{\tau+1}$
\ENDFOR
\STATE Compute advantages $A_\tau$ and rewards $R_\tau$ from the collected trajectories
\STATE Compute total loss $L(\theta, \phi) = L^{CLIP}(\theta) - c_1 L^{VF}(\phi) + c_2 S(\pi_\theta)$
\STATE Store trajectories into replay buffer $\mathcal{D}$
\IF{$\mathcal{D} \geq \text{batch\_size}$}
\FOR{$i = 1$ to $N_{\text{mini\_batch}}$} 
\STATE Sample mini-batch of trajectories from $\mathcal{D}$: $(s_\tau, a_\tau, r_\tau,s_{\tau+1})$

\STATE Update policy and value network parameters $\theta$ and $\phi$ using gradient ascent on $L(\theta, \phi)$
\ENDFOR   
\STATE Update $\theta_{\text{old}} = \theta$
\STATE Clear replay buffer $\mathcal{D}$ 
\ENDIF
\ENDFOR
\end{algorithmic}
\end{algorithm}

\section{Numerical Results}\label{sec:Numerical Results}
In this section, we evaluate the performance of our proposed algorithm through numerical simulations.

\subsection{Evaluation Setup}
We utilized the PyTorch framework to construct a simulation environment and conducted a comprehensive evaluation of the performance of the proposed solution. The network scenario considered in this paper consists of a LEO satellite edge and a GU. In this scenario, the LEO satellites are uniformly distributed on the same orbital plane, with an angular separation of $4^{\circ}$ between adjacent LEO satellites. They orbit around the Earth in a clockwise direction with the same angular velocity $\mathcal V=0.0002^{\circ}/s$. The number of LEO satellites is $J=12$, and their orbital altitude is $780km$, while the radius of the Earth is $6371km$. The GU is located on the Earth's surface and is positioned along a straight line passing through the Earth's center.

In our simulations, we set the number of scheduling periods $T$ to $50$. In each scheduling period $\tau$, $20$ security-sensitive computing tasks are generated, with an average data size $\lambda=20MB$. The average number $\mu$ of malicious satellites within the communication range is $3$. The parameter for channel gain $\beta_{o}$ is $- 37 dB$, the transmission power $P^{Ka}$ of the GU is $5W$, the AWGN power $\hat {\sigma }^{2}$ is $10^{-6}W$, and the total bandwidth $B_{Ka}$ for the Ka-band is $20MHz$. The average number $q^{local}$ of CPU cycles for the GU to process 1 bit locally is $80$, the computing frequency $f^{local}$ of the GU's local device is $6.5GHz$, the average number $q^{en}$ of CPU cycles for the GU to encrypt 1 bit is $20$, the encryption computing frequency $f^{en}$ of the GU is $3.0GHz$, and the parameter $k$ for the hardware architecture of the local device is $10^{-31}W \cdot s^3/cycle^3$. The threshold ${\rho }$ for the success probability of task offloading is set to $70\%$, and the weight factors $\beta_{1}$ and $\beta_{2}$ for the optimization objective are both set to $1$. Table \ref{table:training parameters of PPO} lists the training parameters of PPO.

\begin{table}[!t]  
	\begin{center}
		\centering
		\caption{training parameters of PPO.}  
		\label{table:training parameters of PPO} 
		\footnotesize
		\begin{tabular}{p{5.8cm}p{2cm}} \hline
			\textbf{Parameter} & \textbf{Value}\\
			Total timesteps  & $5\times  10^5$ \\
			Update interval  & $5$ \\
			Batch size  & $64$ \\
			Gamma  & $0.99$ \\
			Gae lambda  & $0.95$ \\
			Clip range  & $0.2$ \\
			\hline
		\end{tabular}   
	\end{center}   
\end{table}

\subsection{Convergence performance}

In this section, we will evaluate the converge performances of the PPO algorithm with different hyperparameters. We also compare it with other DRL algorithms.

\subsubsection{Convergence Performance Under Different Update Interval}

\begin{figure}[!t]
\centering
\includegraphics[width=0.7\columnwidth]{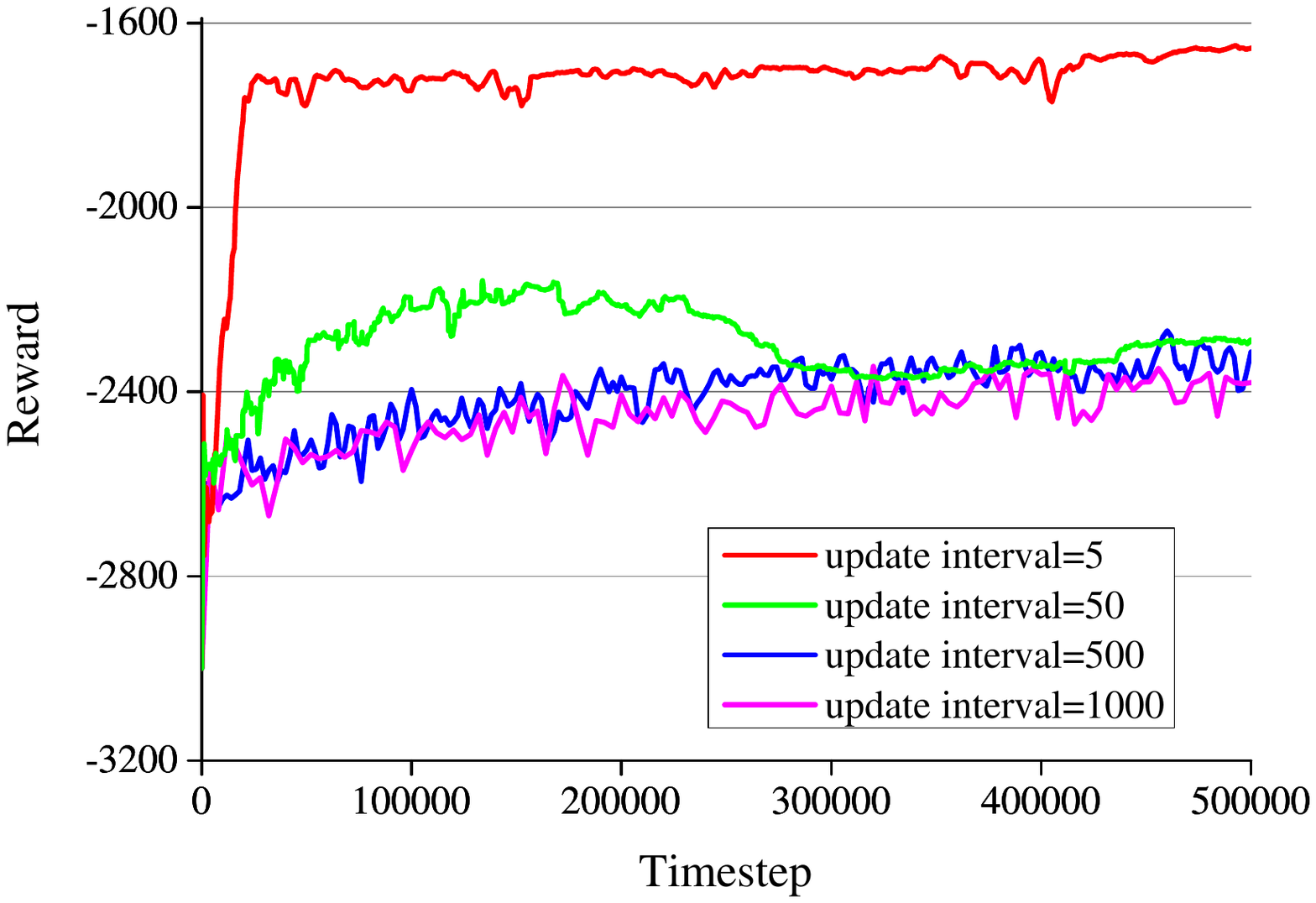}
\caption{Convergence performance under different update interval.}
\label{fig:Result-update interval}
\end{figure}

The \textit{update interval} hyperparameter in training specifies the number of steps to be executed in each environment before performing a model update. To investigate the impact of this hyperparameter on the convergence of the PPO algorithm, we compared the convergence curves of reward values for different update intervals in our experiment. As shown in Figure \ref{fig:Result-update interval}, when the update interval is set to 5, the reward curve reaches a higher value at approximately 30,000 steps, and then maintains a slight upward trend, eventually reaching the highest reward value. This indicates that with a smaller number of steps, the reward can be quickly improved, leading to the attainment of the optimal strategies in a relatively short time. Conversely, when the update interval is set to 50, the reward curve initially increases but starts to exhibit a declining trend at around 150,000 steps. It eventually converges to a lower reward value at approximately 480,000 steps. For update intervals of 500 and 1000, both reward curves show a slow upward trend. However, the final converged reward values are relatively lower. These findings suggest that longer update intervals can enhance model performance to some extent, but there are limitations in reaching the optimal solution. Hence, it is evident that the selection of the update interval parameter significantly affects the algorithm's performance and convergence speed. This analysis aids in selecting an appropriate update interval hyperparameter to achieve improved model performance and results.

\subsubsection{Convergence Performance Under Different Learning Rates}

\begin{figure}[!t]
	\centering
	\includegraphics[width=0.7\columnwidth]{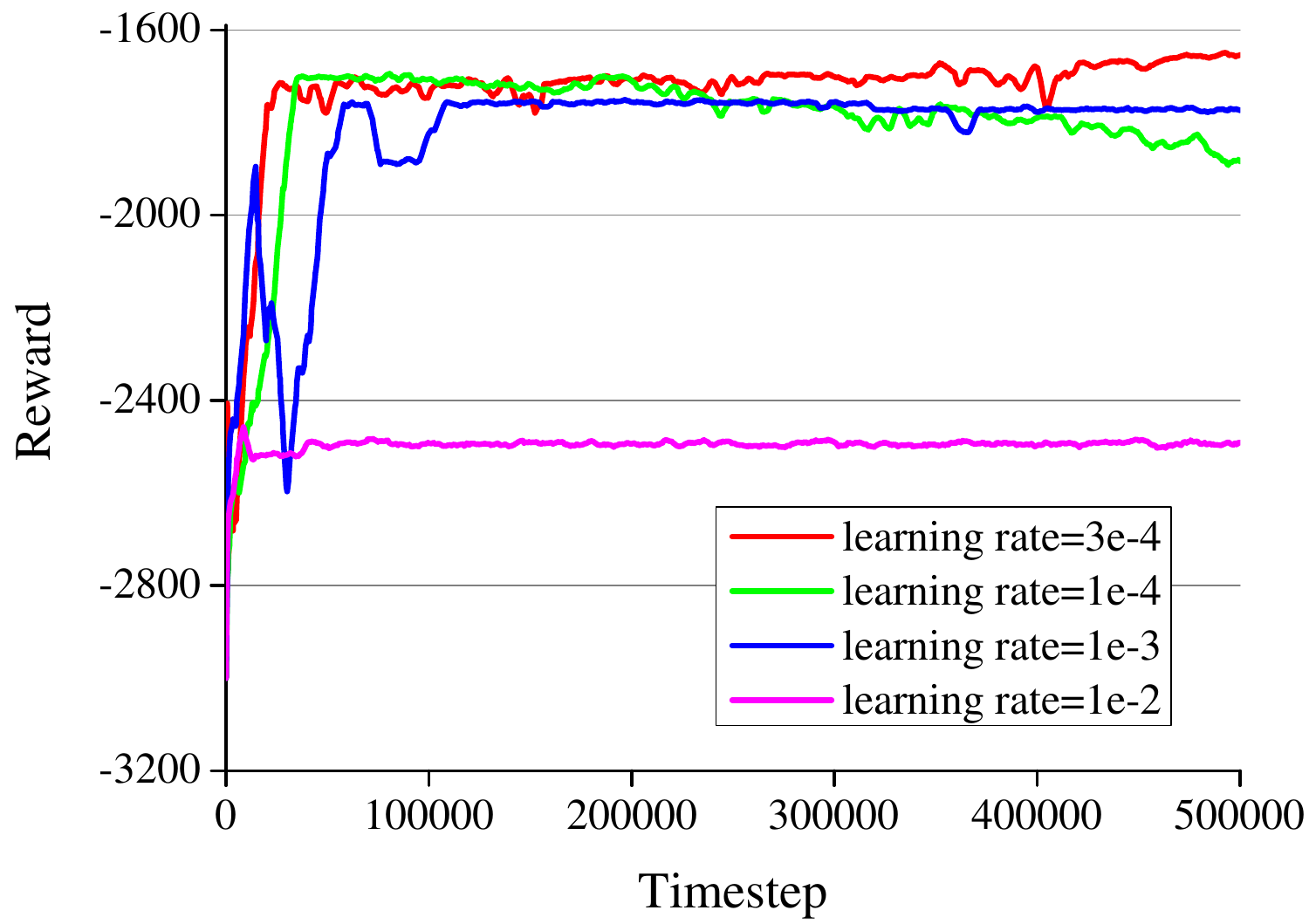}
	\caption{Convergence performance under different learning rates.}
	\label{fig:Result-learning rate}
\end{figure}

To explore the impact of the learning rate on the convergence of the PPO algorithm, we compared the convergence curves of reward values for different learning rates in our experiment. As shown in Figure \ref{fig:Result-learning rate}, when the learning rate is set to $3\times 10^{-4}$, the reward curve converges after approximately 490,000 steps. However, when the learning rate is set to $1\times 10^{-4}$, the curve reaches its peak after about 50,000 steps but then shows a downward trend. This may be due to the learning rate being too low, preventing the model from adequately learning and adjusting its strategies. When the learning rate is set to $1\times 10^{-3}$, the curve converges after approximately 110,000 steps, but the final reward value is lower. This could be because the learning rate is too high, causing the parameter updates to be too large, and the model misses out on better strategies during the training process. Lastly, when the learning rate is set to $1\times 10^{-2}$, the curve quickly converges to a lower reward value shortly after training begins. This indicates that the learning rate is too high, causing the model to skip over better strategies early in training and preventing further performance improvement. The experiment demonstrates that the learning rate significantly affects the convergence speed and performance of our proposed PPO algorithm. Choosing an appropriate learning rate can help the model better fit the training data and find the optimal offloading strategy for the task.

\subsubsection{Convergence Performance Under Different DRL Algorithms}

\begin{figure}[!t]
	\centering
	\includegraphics[width=0.7\columnwidth]{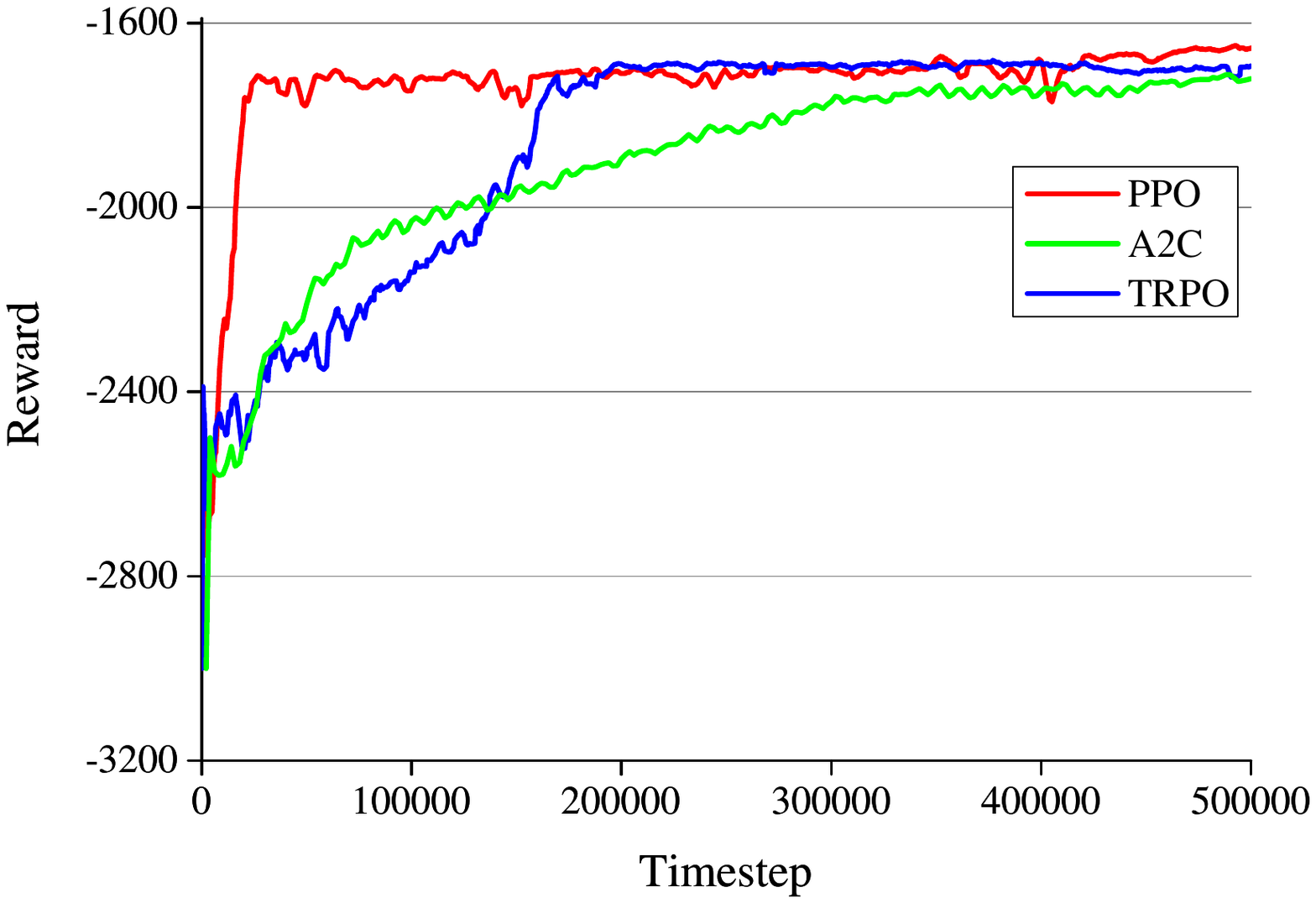}
	\caption{Convergence performance under different DRL algorithms.}
	\label{fig:Result-Different DRL}
\end{figure}

To validate the convergence and adaptability of the PPO algorithm in the ISTN scenario, we compared the reward convergence curves of PPO with those of the Advantage Actor-Critic (A2C) and the Trust Region Policy Optimization (TRPO) algorithms. As shown in Figure \ref{fig:Result-Different DRL}, the reward curve of the PPO algorithm quickly rises and converges shortly after training begins. At the end of training, the PPO algorithm achieves the highest reward value. In contrast, the reward curve of the A2C algorithm exhibits a slower ascent and converges at around 170,000 steps. The curve of the TRPO algorithm also rises at a slower pace and converges towards the end of training. PPO introduces importance sampling ratio clipping, which ensures that the sampled data in the update steps are more effectively utilized, thereby improving sampling efficiency and data utilization. In comparison, A2C and TRPO require more sampled data for policy updates, resulting in relatively lower efficiency. The experiment demonstrates that the PPO algorithm exhibits superior convergence and adaptability in this environment, enabling faster learning of effective offloading strategies.

\subsection{Performance Analysis for Security-Sensitive Task Offloading Algorithm}

To further evaluate the performance of the proposed solution, we included the following algorithms in our experiments for comparison:

\textbf{A2C-based:} This algorithm is based on A2C, aiming to find the optimal task offloading policy. The A2C algorithm combines policy gradient and value function estimation, improving the task offloading policy by maximizing the advantage function.

\textbf{TRPO-based:} This algorithm is based on TRPO, aiming to find the optimal task offloading policy. The TRPO algorithm utilizes gradient ascent to maximize policy performance and ensures that the magnitude of policy updates is constrained by enforcing a limit on the KL divergence of policy updates.

\textbf{Greedy:} During each scheduling period, the algorithm generates 1000 different task offloading strategies randomly and selects the one with the best performance based on the defined performance metrics.

\textbf{Round-Robin:} This algorithm follows a fixed order to sequentially allocate the generated computing tasks in each scheduling period to the local device and each LEO satellite, ensuring that both the local device and each LEO satellite have the opportunity to participate in the processing of computing tasks.

\textbf{All-Local:} This algorithm allocates all computing tasks to the local device for processing without offloading them to LEO satellites. This approach is suitable when the local device has sufficient computational resources and processing capabilities, which helps to avoid communication latency and privacy risks associated with offloading.

\textbf{All-Offloading:} This algorithm offloads all computing tasks to LEO satellites for processing, sequentially allocating tasks to each LEO satellite in order of proximity. This offloading algorithm is suitable for scenarios where the local device has limited computational capabilities and low communication transmission latency, allowing for efficient utilization of computational resources at the LEO satellite edge.

\subsubsection{Total Cost Under Different Task Data Sizes}

\begin{figure} [!t]
	\centering
	\centering
	\includegraphics[width=0.7\columnwidth]{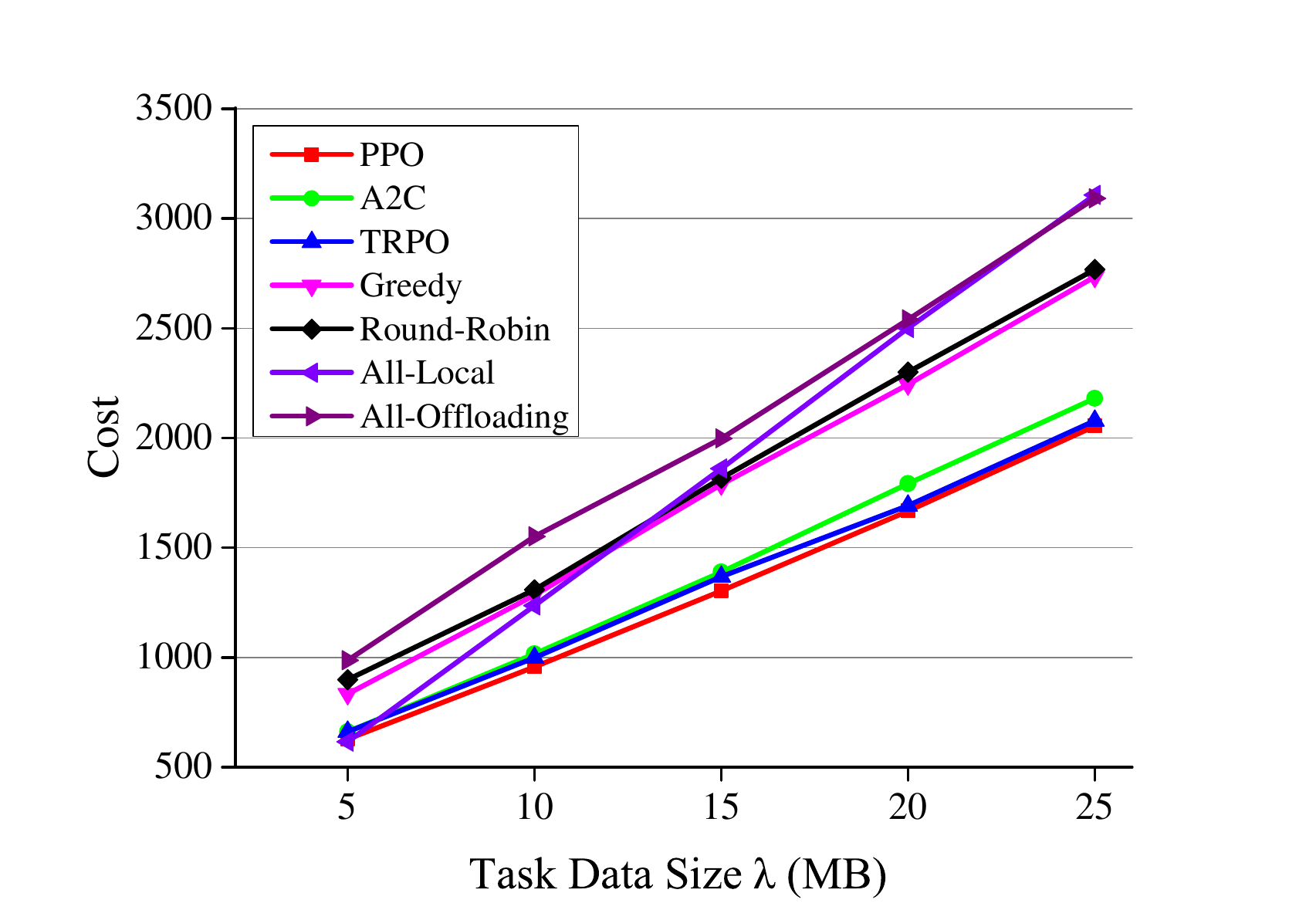}
	\caption{Total cost under different task data sizes.}
	\label{fig:Result-task data size}
\end{figure}

To evaluate the adaptability of the PPO algorithm to different task data sizes, we varied the mean value $\lambda$ of the task data size. The results, as shown in Figure~\ref{fig:Result-task data size}, reveal that the system cost for all algorithms increases with an increase in task data size. This is because larger data sizes result in multiplied delays and energy consumption costs for task offloading. Among the algorithms, the Round-robin scheduling algorithm, Full local algorithm, and Full offloading algorithm yield the highest system costs as the task data size increases. This is because using fixed offloading strategies leads to wastage of computational resources when the system needs to handle larger task demands. Additionally, the Random algorithm exhibits poor adaptability to environmental changes, resulting in higher system costs. Although the A2C-based and TRPO-based algorithms demonstrate lower system costs as the task data size increases, they remain relatively higher compared to the PPO-based algorithm. This could be attributed to the introduction of a clipping function in the PPO-based algorithm, which better controls the magnitude of policy updates. In conclusion, under varying task data sizes, the PPO-based algorithm consistently achieves the lowest system costs, showcasing its outstanding performance in adapting to different task data sizes.

\subsubsection{Total Cost Under Different Local Computing Frequencies}

\begin{figure} [!t]
	\centering
	\includegraphics[width=0.7\columnwidth]{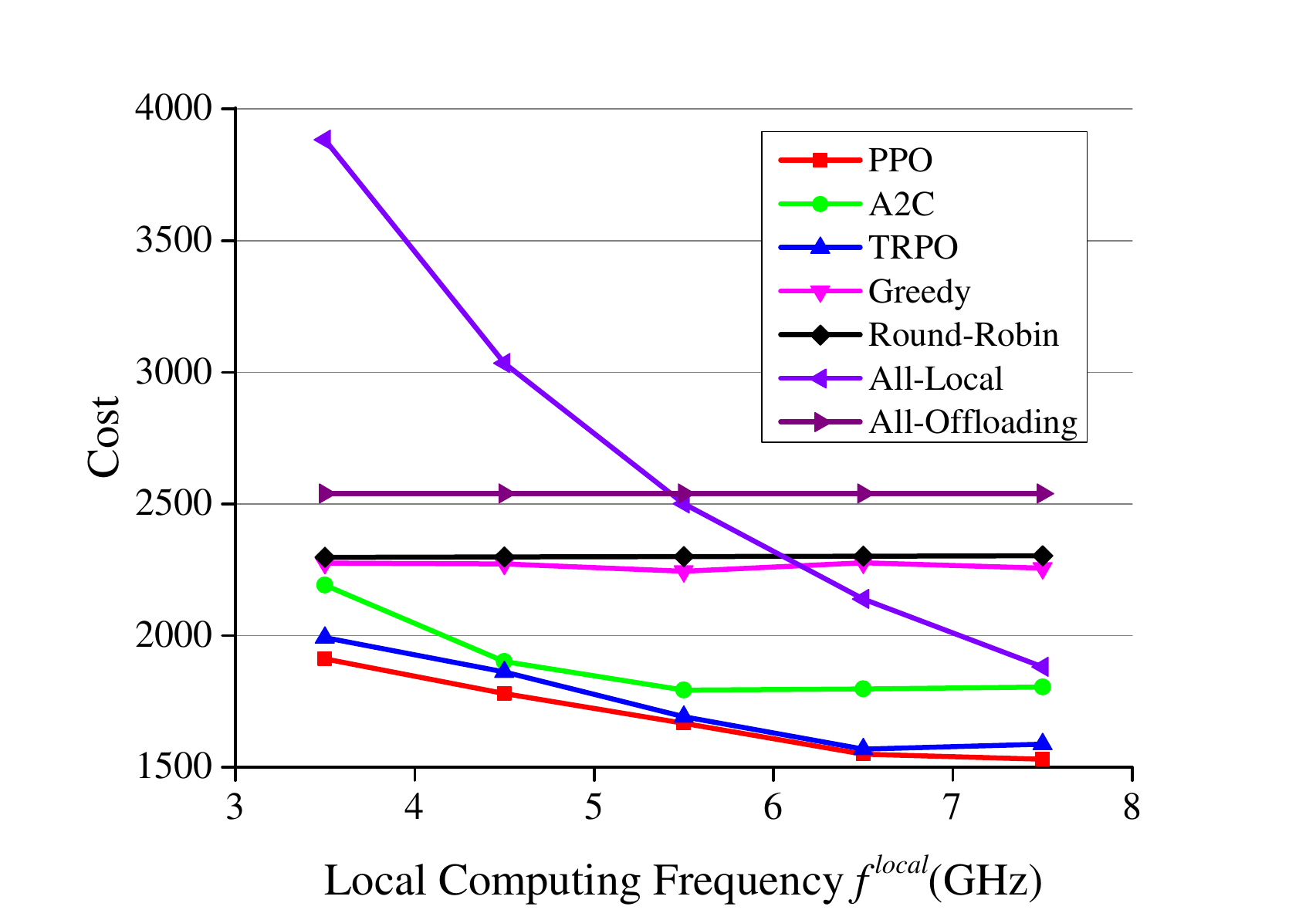}
	\caption{Total cost under different local computing frequencies.}
	\label{fig:Result-local computing frequency}
\end{figure}

We also analyzed the impact of the local device's computational frequency on the system cost, and the results are shown in the Figure \ref{fig:Result-local computing frequency}. Generally, the system cost for most algorithms decreases as the local device's computational frequency increases. This is because the improved computational capability of the local device leads to significant reductions in task computation time. When the local device's computational frequency increases from 3.5 GHz to 5.5 GHz, the system cost of the A2C algorithm gradually decreases. However, when the frequency exceeds 5.5 GHz, the A2C algorithm's system cost does not further decrease significantly. This could be attributed to the relatively lower sampling efficiency and data utilization of the A2C algorithm, limiting its optimization effectiveness when environmental conditions change. At a local device computational frequency of 3.5 GHz, the Full local algorithm yields the highest system cost. This is because, with lower computational capabilities of the local device, the majority of tasks are offloaded to the LEO satellite edge server, resulting in an overloaded LEO satellite and longer task waiting times. Additionally, the increased offloading frequency significantly increases the likelihood of malicious satellite attacks, ultimately leading to higher system costs. As the local device's computational frequency gradually increases, more tasks tend to be retained for local computation, leading to a rapid decrease in system cost for the Full local algorithm. On the other hand, the Round-robin scheduling algorithm, Random algorithm, and Full offloading algorithm exhibit almost no change in system cost as the computational frequency increases. This is because they are static scheduling strategies that do not effectively adapt to environmental changes. Despite achieving lower system costs, the TRPO algorithm remains higher compared to the PPO algorithm. In conclusion, under varying local device computational frequencies, the PPO algorithm consistently achieves the lowest system cost, indicating its ability to learn optimal task offloading strategies.

\subsubsection{Total Cost Under Different Numbers of Malicious Satellites}

\begin{figure} [!t]
	\centering
	\centering
	\includegraphics[width=0.7\columnwidth]{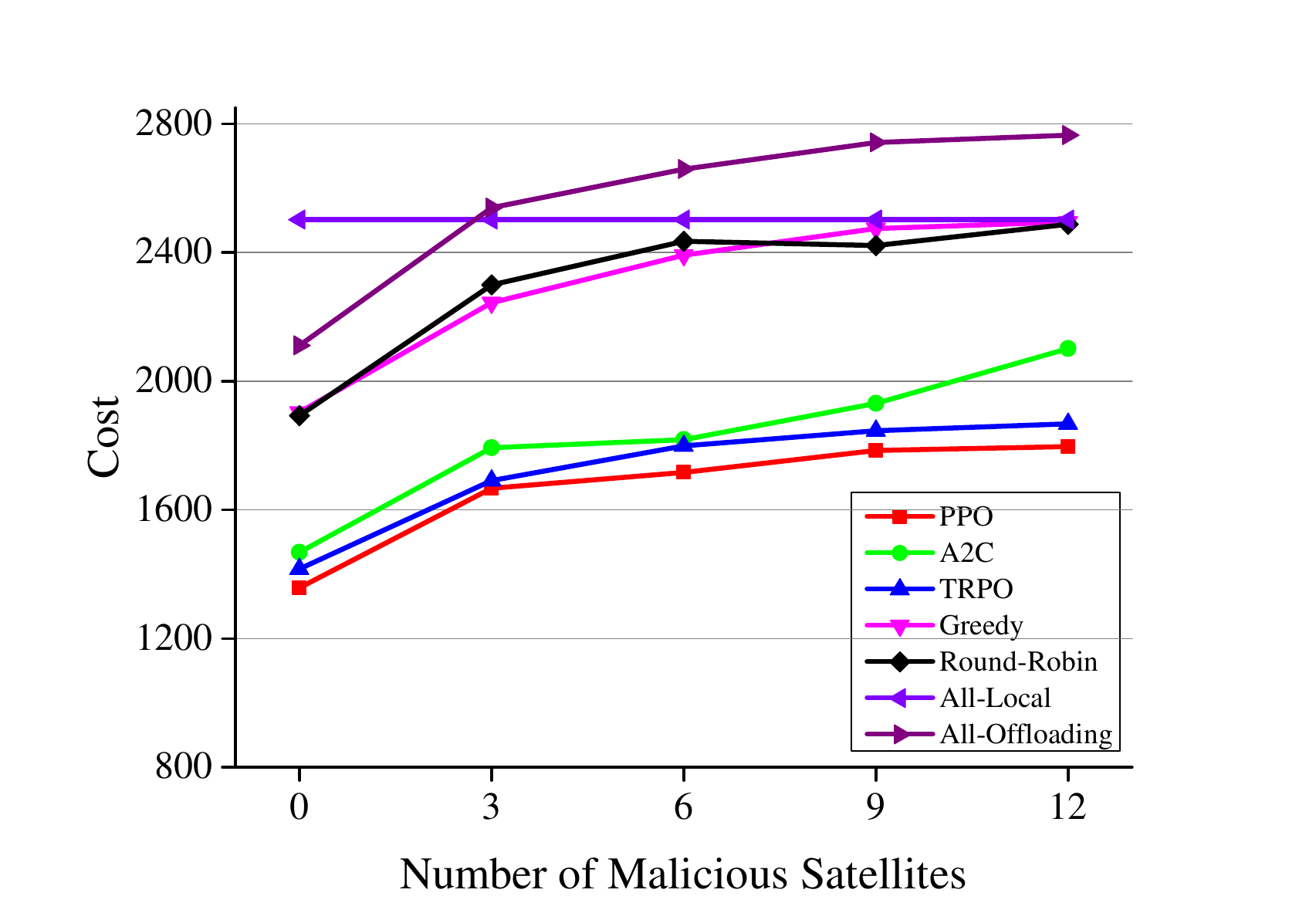}
	\caption{Total cost under different average number of malicious satellites.}
	\label{fig:Result-average number of malicious satellites}
\end{figure}

To investigate the impact of the average number of malicious satellites $\mu$ on the system cost, we adjusted $\mu$ to different values: 0, 3, 6, 9, and 12. The results are shown in Figure \ref{fig:Result-average number of malicious satellites}. As expected, with an increase in the average number of malicious satellites, most algorithms demonstrate varying degrees of increased system cost. This phenomenon is primarily due to the higher probability of task offloading being attacked as the average number of malicious satellites within the communication range increases, leading to an increase in the number of attacks. Additionally, as the average number of malicious satellites increases, the GU tends to keep a larger portion of the tasks for local computation. While this can protect the tasks from attacks, it also results in a heavier workload on the local device, leading to significant increases in queuing time and energy consumption. It is worth noting that the system cost remains unchanged for the All-Local algorithm since it keeps all tasks for local computation, thus avoiding attacks from malicious satellites. Compared to the Greedy algorithm, Round-Robin algorithm, All-Local algorithm, and All-Offloading algorithm, the PPO-based algorithm, A2C-based algorithm, and TRPO-based algorithm, which employ DRL, achieve lower system costs. Particularly, the PPO-based algorithm achieves the lowest system cost, indicating that it can allocate tasks more effectively when facing different average numbers of malicious satellites, and strike a balance between attack risk and system performance.

\section{Conclusion}\label{sec:conclusion}
In this paper, we explore security-sensitive task offloading schemes in integrated satellite-terrestrial networks, aiming to enhance the performance and security of ground user offloading by leveraging LEO satellite edge computing. Our objective is to minimize user offloading delay, energy consumption, and the number of attacks while satisfying reliability constraints. To achieve this, we propose a PPO-based algorithm for optimizing task offloading strategies, which jointly optimizes task allocation decisions and offloading orders. Simulation results demonstrate that our proposed algorithm, through agent training, achieves the lowest system cost compared to other algorithms and learns optimal task offloading strategies. This research holds significant importance for the application of satellite edge computing, as it allows for the efficient utilization of LEO satellite edge computing resources to provide faster, more efficient, and secure task offloading services for ground users.

\section*{Acknowledgments}	
This work is supported by National Natural Science Foundation of China (No. 61802383), Research Project of Pazhou Lab for Excellent Young Scholars (No. PZL2021KF0024), and Guangzhou Basic and Applied Basic Research Foundation (No. 202201010330, 202201020162).

\begin{IEEEbiographynophoto}{Wenjun Lan}
received his Bachelor degree in computer science and technology from Wuyi University in 2021. He is currently pursuing his Master degree at Guangzhou University, China. His research interests include machine learning and mobile edge computing.
\end{IEEEbiographynophoto}

\begin{IEEEbiographynophoto}{Kongyang Chen}
is an Associate Professor at Guangzhou University, China. He received his PhD degree in computer science from the University of Chinese Academy of Sciences, China. His research interests are artificial intelligence, edge computing, blockchain, IoT, etc.
\end{IEEEbiographynophoto}

\begin{IEEEbiographynophoto}{Jiannong~Cao}
is currently the Otto Poon Charitable Foundation Professor in Data Science and the Chair Professor of Distributed and Mobile Computing in the Department of Computing at The Hong Kong Polytechnic University (PolyU), Hong Kong. He is also the Dean of Graduate School, the director of Research Institute for Artificial Intelligence of Things (RIAIoT) in PolyU, the director of the Internet and Mobile Computing Lab (IMCL). His research interests include distributed systems and blockchain, wireless sensing and networking, big data and machine learning, and mobile cloud and edge computing. He has served the Chair of the Technical Committee on Distributed Computing of IEEE Computer Society 2012-2014, a member of IEEE Fellows Evaluation Committee of the Computer Society and the Reliability Society, a member of IEEE Computer Society Education Awards Selection Committee, a member of IEEE Communications Society Awards Committee, and a member of Steering Committee of IEEE Transactions on Mobile Computing. He has also served as chairs and members of organizing and technical committees of many international conferences, including IEEE INFOCOM, IEEE PERCOM, IEEE IoTDI, IEEE ICPADS, IEEE CLOUDCOM, SRDS and OPODIS, and as associate editor and member of the editorial boards of many international journals, including IEEE TC, IEEE TPDS, IEEE TBD, IEEE IoT Journal, ACM ToSN, ACM TIST, ACM TCPS.
He is a member of Academia Europaea, a fellow of the Hong Kong Academy of Engineering Science, a fellow of IEEE, a fellow of China Computer Federation (CCF) and an ACM distinguished member.
\end{IEEEbiographynophoto}

\begin{IEEEbiographynophoto}{Yikai Li}
is a master student at Guangzhou University. His research interests are mobile edge computing and federated learning.
\end{IEEEbiographynophoto}

\begin{IEEEbiographynophoto}{Ning Li}
received the Ph.D. degree from the National Engineering Research Center for Multimedia Software, Wuhan University, China, in 2022. He is currently conducting post-doctoral research with the Institute of Artificial Intelligence, Guangzhou University, China. His research interests include artificial intelligence, computer vision, and data analysis.
\end{IEEEbiographynophoto}

\begin{IEEEbiographynophoto}{Qi Chen}
received the Ph.D. degree in mathematics from the Guangzhou university, China in 2011. Since 2017, he has been with Guangzhou University. His research interests include cryptography, coding theory and Blockchain.
\end{IEEEbiographynophoto}

\begin{IEEEbiographynophoto}{Yuvraj Sahni}
received the PhD degree from The Hong Kong Polytechnic University, Hong Kong, in 2021. He is currently a Research Assistant Professor at The Hong Kong Polytechnic University, Hong Kong. His research interests include edge computing, IoT, and smart buildings.
\end{IEEEbiographynophoto}

\end{document}